\documentclass[preprint]{emulateapj}

\begin{document}

\title{The Mass-Radius(-Rotation?) Relation for Low-Mass Stars}

\author{
 Adam L. Kraus\altaffilmark{1},
 Roy A. Tucker\altaffilmark{2},
 Michael I. Thompson\altaffilmark{3},
 Eric R. Craine\altaffilmark{4},
 Lynne A. Hillenbrand\altaffilmark{5}
 }

 \altaffiltext{1}{alk@ifa.hawaii.edu; Hubble Fellow, University of 
Hawaii-IfA, 2680 Woodlawn Dr, Honolulu, HI 96822}
 \altaffiltext{2}{Goodricke-Pigott Observatory, 5500 West Nebraska Street, 
Tucson, AZ 85757}
 \altaffiltext{3}{Global Network of Astronomical Telescopes, Inc., Tucson, 
AZ}
 \altaffiltext{4}{Western Research Company, Inc., 3275 W. Ina Rd., Suite 
217, Tucson, AZ 85741; Global Network of Astronomical Telescopes, Inc., 
Tucson, AZ; Dept. of Physics, Colorado State University, Fort Collins, 
CO}
 \altaffiltext{5}{Dept. of Astronomy MC 249-17, California 
Institute of Technology, 1200 E. California Blvd., Pasadena, CA 91125}

\begin{abstract}

The fundamental properties of low-mass stars are not as well understood as 
those of their more massive counterparts. The best method for constraining 
these properties, especially masses and radii, is to study eclipsing 
binary systems, but only a small number of late-type ($\ge$M0) systems 
have been identified and well-characterized to date. We present the 
discovery and characterization of six new M dwarf eclipsing binary 
systems. The twelve stars in these eclipsing systems have masses spanning 
0.38-0.59 $M_{\sun}$ and orbital periods of 0.6--1.7 days, with typical 
uncertainties of $\sim$0.3\% in mass and $\sim$0.5-2.0\% in radius. 
Combined with six known systems with high-precision measurements, our 
results reveal an intriguing trend in the low-mass regime. For stars with 
$M=$0.35-0.80 $M_{\sun}$, components in short-period binary systems 
($P\la$1 day; 12 stars) have radii which are inflated by up to 10\% 
($\mu=4.8\pm1.0\%$) with respect to evolutionary models for low-mass 
main-sequence stars, whereas components in longer-period systems ($>$1.5 
days; 12 stars) tend to have smaller radii ($\mu=1.7\pm0.7\%$). This trend 
supports the hypothesis that short-period systems are inflated by the 
influence of the close companion, most likely because they are tidally 
locked into very high rotation speeds that enhance activity and inhibit 
convection. In summary, very close binary systems are not representative 
of typical M dwarfs, but our results for longer-period systems indicate 
the evolutionary models are broadly valid in the $M\sim$0.35-0.80 
$M_{\odot}$ regime.

\end{abstract}


\section{Introduction}

M dwarfs are ubiquitous in the solar neighborhood and constitute the 
majority of the stellar content in our galaxy, but their fundamental 
properties are not as well understood as those of their more massive 
brethren. These measurements are crucial for calibrating stellar 
evolutionary models, inferring accurate masses and radii for transiting 
exoplanets, and understanding the evolution of low-mass companions in 
compact binaries. These properties (mass, radius, luminosity, and 
effective temperature) are typically calibrated by observations of binary 
systems. However, since M dwarfs are intrinsically faint, a very limited 
sample of M dwarf eclipsing binaries (MDEBs) is accessible and suitable 
for more detailed study. The radii of low-mass stars have been 
particularly difficult to study since they can be measured with high 
precision ($\sigma \la 1-2 \%$) only in double-lined eclipsing binary 
systems, which must be identified in wide-field, multi-epoch variability 
studies. To date, only $\sim$10 systems with primary masses $\la$0.6 
$M_{\sun}$ have been identified, and only a handful of those have been 
characterized with the necessary precision.

Preliminary results for the few well-studied systems have found a 
troubling level of disagreement with theoretical models. Most of the 
components in low-mass eclipsing systems appear to be 5--15\% larger than 
theoretical models would predict (e.g. Lacy 1977; Leung \& Scheider 1978; 
Lopez-Morales \& Ribas 2005; Bayless \& Orosz 2006; Irwin et al. 2009), 
and this excess is seen for a wide range of stellar masses (0.2--0.8 
$M_{\odot}$). However, it is unclear whether the larger radii indicate a 
general problem in models (i.e. due to missing opacities) or a systematic 
effect specific to eclipsing systems. For example, most close binary 
systems are tidally locked into very rapid rotation, which could lead to 
stronger magnetic field interactions than for slow-rotating single stars 
(e.g. Chabrier et al. 2007) that inhibit the efficiency of convection in 
the stellar envelopes. This explanation seems especially plausible given 
that the closest binary systems are known to show extensive spot coverage 
(Morales et al. 2009; Windmiller et al. 2010) and strong H$\alpha$ 
emission, both of which can be signs of chromospheric activity and strong 
magnetic fields; similar effects are also seen for young low-mass binaries 
(e.g. Stassun et al. 2006). These active stars could also yield incorrect 
radius measurements if the spots are not randomly distributed; Morales et 
al. (2010) have suggested that concentration of spots near the poles could 
also explain the discrepancy in measured radii. The most straightforward 
test of these hypotheses would be to characterize longer-period systems 
and determine if the mass-radius relation depends on the orbital period 
(and hence the rotation of the stars). However, this test would require a 
much larger sample of systems, and is hampered because long-period systems 
are less likely to eclipse and have a lower eclipse duty cycle (and hence 
are more difficult to identify and study).

The few known MDEB systems been discovered serendipitously in programs 
such the OGLE microlensing survey and the TReS transiting exoplanet 
survey. There have been searches for MDEBs in existing wide-field 
variability surveys like ROTSE (Akerlof et al. 2003) and ASAS (Pojmanski 
et al. 2005), but only one new MDEB has been reported (GU Boo;  
Lopez-Morales \& Ribas 2005) since the shallow depth of these surveys 
($V_{lim}\la$13) limits their sensitivity to intrinsically faint variable 
stars ($M_V=9$ for an M0 dwarf, $M_V=13$ for an M6 dwarf). Any survey to 
identify a significant number of new MDEBs must extend significantly 
fainter than current-generation systems while still studying a significant 
fraction of the sky. To this end, we have launched a program to identify 
and characterize new MDEBs in deeper variability surveys that are now 
being released, beginning with the 1st MOTESS-GNAT survey (MG1; Kraus et 
al. 2007).

In this paper, we describe the first six M dwarf eclipsing binary systems 
to emerge from our search. In Section 2, we briefly outline the discovery 
of these systems. In Sections 3 and 4, we describe the photometric and 
spectroscopic observations that contributed to the discovery and analysis 
of these systems, while in Section 5, we present the analysis that 
ultimately yields precise masses and radii for the components of each 
system. Finally, in Section 6, we discuss some of the implications of our 
updated mass-radius relation for stellar evolutionary models, and we 
specifically discuss the potential role that stellar rotation plays in 
determining the mass-radius relation for low-mass stars.

\section{Discovery}

 \begin{deluxetable*}{lccrrrrrrr}
 \tabletypesize{\scriptsize}
 \tablewidth{0pt}
 \tablecaption{New M-Dwarf Eclipsing Binaries}
 \tablehead{\colhead{Name} & \colhead{RA} & \colhead{DEC} & 
 \colhead{$R_{MG1}$} & \colhead{$K_s$} & \colhead{$m_{bol}$} & 
\colhead{$\mu_{\alpha}$} & \colhead{$\mu_{\delta}$} & 
\colhead{$\sigma_{\mu}$} & \colhead{SpT}
 \\ 
 \colhead{} & \multicolumn{2}{c}{(J2000)} & \colhead{(mag)} & 
\colhead{(mag)} & \colhead{(mag)} & \multicolumn{3}{c}{(mas yr$^{-1}$)} & 
\colhead{}
 }
 \startdata
MG1-78457&03 26 20.7&+03 12 36&16.3&12.69&15.38&5&$-$4&4&M3.3$\pm$0.4\\
MG1-116309&04 48 09.6&+03 17 47&14.9&11.91&14.31&16&10&5&K7.9$\pm$0.4\\
MG1-506664&07 43 11.5&+03 16 22&14.7&11.85&14.35&9&$-$12&4&M1.0$\pm$0.5\\
MG1-646680&10 30 55.3&+03 34 27&16.0&13.30&15.78&$-$24&$-$21&3&M1.0$\pm$0.1\\
MG1-1819499&20 11 51.4&+03 37 20&15.0&12.13&14.63&$-$18&$-$33&4&M1.1$\pm$0.5\\
MG1-2056316&23 14 38.3&+03 39 52&14.8&11.64&14.25&$-$42&$-$65&4&M2.6$\pm$1.1\\
 \enddata
 \tablecomments{The derived properties ($m_{bol}$, $\mu$, and SpT) 
were calculated using the multi-catalog data mining procedure we described 
in Kraus \& Hillenbrand (2007). The photometric uncertainties are 
$\sim$0.1--0.2 mag for MG1 $R$, $\sim$0.02 mag for 2MASS $K_s$, and 
$\sim$0.05 mag for $m_{bol}$. The $R$ magnitude uncertainty is systematic 
since it is calibrated into the less well-defined USNOB1.0 R magnitude 
system. The uncertainty $\sigma_{\mu}$ is the uncertainty along each 
axis.}
 \end{deluxetable*}

All six systems were identified in the variability catalog of the First 
MOTESS-GNAT variable star survey (MG1; Kraus et al. 2007). MG1 is a deep, 
wide-field imaging survey which was conducted with the Moving Object and 
Transient Event Search System (MOTESS; Tucker 2007). MOTESS is composed of 
three 14-inch telescopes which operate in drift-scan mode to conduct deep 
multi-epoch imaging near the celestial equator. The MG1 survey was 
compiled from observations taken during the first two years of MOTESS 
operation and covers a total field of 300 deg$^2$, with observations of 
$\sim$100-120 deg$^2$ taken twice each night. A total of $\sim$1.6 million 
sources were observed at $\sim$150-250 epochs in this campaign; the MG1 
survey identified 26042 of them to be variable star candidates using the 
Welch-Stetson variability test (Welch \& Stetson 1993).

The observing cadence of MG1 (twice per night for two observing seasons) 
was too sparse to allow for the identification of low-mass EBs by their 
light curves alone.  There are typically 150-250 observations for each 
source in MG1, so only $\sim$5-20 observations over an interval of two 
years will have occurred during an eclipse. Another potential complication 
is that the observations occured at intervals of exactly 1 sidereal day, 
so variability on shorter timescales will be subject to aliasing.

We addressed these issues by disregarding periodicity and light curve 
morphology in favor of a more basic diagnostic of possible eclipses:  the 
presence of an excess of faint observations, as determined by the skew of 
the brightness distribution. This criterion could be biased against the 
detection of extremely short-period systems (where the eclipse duty cycle 
is $\ga$50\%) because those light curves tend to resemble a balanced 
sinusoidal shape. However, it is very sensitive to long-period systems 
(which are otherwise hardest to identify) because their brightness 
distribution consists of a well-defined Gaussian shape with highly 
significant outliers. MG1 contains 6061 stars that have light curve skews 
of $\ge$1, so we narrowed our search to this subset. We then 
cross-referenced this list of candidates with 2MASS to construct 
($R-K$,$J-K$) and ($J-H$,$H-K$) color-color diagrams and selected the 201 
candidates with colors consistent with the low-mass main sequence. We 
further removed all objects with galactic latitude $|b|<10^o$ (to avoid 
reddened early-type EBs) and visually inspected the remaining curves to 
remove light curves affected by erroneous measurements, leaving a total of 
127 candidates.

As we describe below, we obtained low-resolution optical spectra of these 
candidates to distinguish true M dwarfs from reddened early-type stars, 
yielding $\sim$30 M dwarfs which were likely eclipsing binaries. Finally, 
we performed intensive photometric monitoring of each system with small 
telescopes to confirm its eclipsing nature and establish its period in 
preparation for detailed followup with large-aperture telescopes. As we 
will report in future publications, we have confirmed at least $\sim$20 
new systems; followup for the rest of these systems is ongoing.

We list our newly-discovered MDEB systems in Table 1, along with 
photometry obtained from the discovery survey and from 2MASS (Skrutskie et 
al. 2006). We also list the proper motion, spectrophotometric distance, 
bolometric magnitude, and best-fit spectral type as inferred with the 
astrometric and photometric analysis pipeline described in Kraus \& 
Hillenbrand (2007) and Kraus et al. (in prep). This pipeline uses archival 
astrometry from SDSS, 2MASS, USNO-B1.0, and DENIS to measure the proper 
motion of a source, then uses the corresponding multi-color photometry to 
estimate the best-fit spectral type and spectrophotometric distance 
against a grid of standard spectral energy distributions (SEDs).

\section{Photometric Observations}

\begin{deluxetable}{lrrrr}
\tabletypesize{\scriptsize}
\tablewidth{0pt}
\tablecaption{Eclipse Timing Observations}
\tablehead{\colhead{} & \multicolumn{2}{c}{MG1 Data} &
\multicolumn{2}{c}{Followup Data}
\\
\colhead{MG1-} & 
\colhead{$N$} & \colhead{$\sigma$(mag)} &
\colhead{$N$} & \colhead{$\sigma$(mag)}
}
\startdata
78457&168&0.047&291&0.029\\
116309&185&0.040&289&0.033\\
506664&192&0.033&153&0.012\\
646680&222&0.039&334&0.029\\
1819499&181&0.028&220&0.025\\
2056316&120&0.044&876&0.021\\
\enddata
\end{deluxetable}

\begin{deluxetable*}{rcccccccccccccccc}
\tabletypesize{\scriptsize}
\tablewidth{0pt}
\tablecaption{Multicolor Eclipse Observations}
\tablehead{\colhead{} & \multicolumn{6}{c}{Primary Eclipse} &
\multicolumn{6}{c}{Secondary Eclipse}
\\
\colhead{MG1-} & 
\colhead{$N_{obs}$} & \colhead{Epoch} & \colhead{Duration} & 
\colhead{$\sigma_{I}$} & \colhead{$\sigma_{R}$} & \colhead{$\sigma_{V}$} & 
\colhead{$N_{obs}$} & \colhead{Epoch} & \colhead{Duration} & 
\colhead{$\sigma_{I}$} & \colhead{$\sigma_{R}$} & \colhead{$\sigma_{V}$} & 
\\
\colhead{} & \colhead{} & \colhead{(JD-2450000)} & \colhead{(hours)} & 
\colhead{(mag)} & \colhead{(mag)} & \colhead{(mag)} & 
\colhead{} & \colhead{(JD-2450000)} & \colhead{(hours)} &
\colhead{(mag)} & \colhead{(mag)} & \colhead{(mag)} & 
}
\startdata
78457&90&4758.92&3.05&0.009&0.016&0.029&94&4781.92&3.09&0.013&0.039&0.077\\
116309&72&4783.81&3.08&0.016&0.015&0.028&53&4547.66&1.68&0.008&0.008&0.011\\
506664&75&4573.73&2.64&0.007&0.009&0.013&68&4580.70&2.17&0.009&0.010&0.018\\
646680&76&4547.83&2.44&0.013&0.025&0.053&78&4579.77&2.49&0.021&0.049&0.071\\
1819499&80&4738.75&3.01&0.011&0.018&0.014&89&4739.69&3.09&0.006&0.008&0.013\\
2056316&94&4730.79&3.03&0.006&0.009&0.013&93&4755.77&2.95&0.006&0.015&0.012\\
 \enddata
 \tablecomments{The photometric uncertainties $\sigma$ for each 
observation were estimated from the scatter in the observations taken 
before and/or after each eclipse.}
 \end{deluxetable*}

\begin{deluxetable}{lrrrr}
\tabletypesize{\scriptsize}
\tablewidth{0pt}
\tablecaption{Multicolor Eclipse Data}
\tablehead{\colhead{MG1-} & \colhead{Filter} & \colhead{HJD$-$} & 
\colhead{$\Delta M$\tablenotemark{a}} & \colhead{$\sigma_M$}\\
\colhead{} & \colhead{} & \colhead{2450000} & \colhead{(mag)} & 
\colhead{(mag)}
}
\startdata
  78457 & I &  4758.876471 & $-$0.008 & 0.010 \\
  78457 & I &  4758.877802 &  0.001 & 0.009 \\
  78457 & I &  4758.879133 &  0.005 & 0.009 \\
  78457 & I &  4758.880463 &  0.008 & 0.014 \\
  78457 & I &  4758.881794 &  0.020 & 0.009 \\
  78457 & I &  4758.883124 &  0.039 & 0.010 \\
  78457 & I &  4758.884458 &  0.052 & 0.010 \\
  78457 & I &  4758.885789 &  0.065 & 0.010 \\
  78457 & I &  4758.887120 &  0.068 & 0.009 \\
  78457 & I &  4758.888451 &  0.067 & 0.009 \\
  78457 & I &  4758.889782 &  0.107 & 0.012 \\
  78457 & I &  4758.891114 &  0.112 & 0.009 \\
  78457 & I &  4758.892444 &  0.131 & 0.011 \\
  78457 & I &  4758.893775 &  0.176 & 0.010 \\
  78457 & I &  4758.895107 &  0.188 & 0.012 \\
  78457 & I &  4758.896438 &  0.205 & 0.009 \\
  78457 & I &  4758.897769 &  0.233 & 0.010 \\
  78457 & I &  4758.899102 &  0.240 & 0.010 \\
  78457 & I &  4758.900434 &  0.257 & 0.017 \\
  78457 & I &  4758.901764 &  0.277 & 0.013 \\
  78457 & I &  4758.903094 &  0.304 & 0.014 \\
  78457 & I &  4758.904427 &  0.313 & 0.011 \\
  78457 & I &  4758.905759 &  0.339 & 0.009 \\
  78457 & I &  4758.907091 &  0.358 & 0.013 \\
 \enddata
 \tablecomments{Table 4 is published in its entirety in the electronic 
edition of the Astrophysical Journal. A portion is shown here for guidance 
regarding its form and content.}
 \tablenotetext{a}{All differential magnitudes are reported with respect 
to the average of all measurements taken outside of the eclipse, such that 
the magnitude outside of eclipse is $M=0$.}
 \tablenotetext{b}{The photometric uncertainties for each observation 
correspond to the photon counting statistics for the source and 
background. In our light curve fits, we replaced this measurement with the 
observed scatter outside of eclipse (Table 3) if the observed scatter was 
larger, indicating that other noise sources dominated.}
 \end{deluxetable}

\subsection{MG1 Photometry}

The MG1 variable star catalog was produced using an automated pipeline 
that runs in IRAF\footnote{IRAF is distributed by NOAO, which is operated 
by AURA under cooperative agreement with the NSF.}. Dark subtraction and 
flat fielding were performed with standard IRAF tasks, and then aperture 
photometry was measured using the IRAF task QDPhot (Mighell 2000). QDPhot 
is designed to perform fast photometric analysis for data mining of image 
archives and is optimized to minimize runtime while still delivering 
acceptable accuracy and completeness. The primary optimizations are to 
round the stellar centroid to the nearest pixel and to use only a fixed 
pattern of whole pixels in the aperture; these choices result in a small 
increase in uncertainty ($\sim$1-2\%) since the aperture can be offset 
from the stellar centroid by up to 0.7 pixels.

Differential photometry for each source in MG1 was computed with a 
modified implementation of inhomogeneous-ensemble differential photometry 
(IEDP; Honeycutt 1992). In IEDP, all objects are assigned standard 
magnitudes based on the observations that are taken at darktime with the 
best seeing and atmospheric transparency. Since all objects have standard 
magnitudes, they can then all be treated as potential ensemble members. 
The magnitude offset between the nightly instrumental magnitude and the 
standard magnitude of each object is then determined by the mean of the 
difference for an arbitrarily sized ensemble of all objects surrounding 
it. The internal photometric accuracy for each light curve has been 
measured to be $\sim$0.02 mag for the brightest stars ($R\sim13-14$) and 
$\sim$0.04 mag at $R=16$, which is more than sufficient for the purpose of 
eclipsing binary discovery.

All observations reported in MG1 were conducted without a filter to 
maximize sensitivity, so calibration to a standard photometric system is 
not easily achieved. We observed that the detector response is well 
matched by a red photographic plate, so we addressed this challenge by 
calibrating our final photometric results using the $R$ band photometry of 
the USNO-A2.0 catalog. The detector response is not a perfect match, which 
suggests that a small color term is required, plus the original USNO-A2.0 
photographic magnitudes are systematically uncertain by $\sim$0.25 mag, so 
the photometric calibration should be treated with some caution.

In Table 2, we list the total number of MG1 observations for each object 
and the standard deviation of all observations obtained outside eclipse. 
The scatter in the light curve is far less than the typical eclipse 
amplitude, so the eclipse epochs are easily identified. The total number 
of measurements ($\sim$170-270) was large enough to detect eclipses at 
5-20 epochs, but we found that aliasing significantly compromised our 
effort to measure periods. We therefore decided to obtain additional 
high-cadence followup photometry for our candidate eclipsing systems.

\subsection{Followup Photometric Monitoring}

All of our new MDEBs are relatively bright ($R\sim15-16$), so we opted to 
pursue high-cadence followup photometry using 3 small telescopes that 
could be dedicated to the effort for extended periods of time. Each 
eclipsing system was observed in an extended campaign until we detected a 
sufficient number of well-characterized eclipses; we then combined this 
new data with the existing MG1 data to obtain the additional accuracy 
afforded by a 6-8 year time baseline. All observations were obtained 
without filters to maximize sensitivity and because eclipse morphology 
only changes modestly with color. Systems with periods of $<$1 day 
typically yielded the first eclipse within $\la$2--3 days and a full 
period determination within $<$1 week. Systems with periods of 1.5--2.0 
days required a longer observing sequence, but also were typically 
characterized within $<$1 week. As we will report in a future publication, 
we obtained a much longer time series for MG1-2056316 because it is 
serendipitously located in the same field as another MDEB with a much 
longer period.

The first monitoring system, colocated with MOTESS outside Tucson and 
operated by R. Tucker, consists of a Celestron 8 telescope with an Edward 
Byers worm-gear driven mounting and an SBIG ST9 imaging camera. This 
system is manually-operated but capable of unattended tracking and imaging 
of a field of interest all night. Data acquisition, processing of the 
collected images, and photometric reduction was accomplished with Maxim 
DL.

The other two monitoring systems are operated by M. Thompson. The first 
system, located in New Mexico, is a 16" RC Optical Systems 
Ritchey-Chretien OTA on a Bisque Paramount ME robotic mount and uses an 
SBIG STL-6303E camera. The second system, located in Northern California, 
is a 14" Meade SCT OTA on another Bisque Paramount ME mount and uses an 
SBIG STL-1301E camera. These systems are fully automated using a 
combination of custom-written software and the packages CCDsoft and 
TheSky6, both by Bisque. All data were processed using a mix of 
custom-written software and Maxim DL for image calibration, plus Mira Pro 
for aperture photometry.

As for the discovery observations from MG1, we list the total number of 
followup observations for each target and the standard deviation of all 
non-eclipse observations in Table 2.

\subsection{Multicolor Eclipse Photometry}

The geometry of an eclipsing binary system is simple, so approximate 
values for the component temperatures and radii can be obtained from basic 
light curve properties such as the eclipse duration or the primary and 
secondary eclipse depths. However, precise estimation of radii and 
temperatures requires detailed modeling to account for the inclination and 
limb darkening. Our discovery and timing observations were obtained with 
small telescopes using unfiltered data, so those data do not have 
sufficient precision, time cadence, or color information to serve this 
purpose. We instead addressed this requirement by obtaining updated light 
curves of each system's primary and secondary eclipse using the 
roboticized Palomar 60" telescope (P60; Cenko et al. 2006).

The P60 operates solely as an optical imager and is controlled by an 
automated queue-scheduling routine. Its camera has an FOV of 11\arcmin\, 
and a pixel scale of 0.378\arcsec\, pix$^{-1}$, but we operated in a 
5.5x11\arcmin\, subarray in order to reduce the read time to $\sim$15s, 
matching our very short exposures. Each monitoring window was set 1 hour 
wider than the eclipse to obtain constant (non-eclipse) observations and 
to allow for some flexibility in the queue-scheduling software. However, 
our observations were sometimes interrupted by weather or higher-priority 
events, so some observations were truncated and do not include 
post-eclipse brightness measurements. Our observation sequence used 
alternating exposures with $VRI$ filters to obtain coeval 3-color light 
curves. In all cases, we used exposure times of 30s in $V$ and 15s in $R$ 
and $I$, so the interval between subsequent observations in the same 
filter is ~107 sec.

All images from the P60 are automatically bias-subtracted and flat-fielded 
as part of the data acquisition pipeline. We extracted magnitudes for all 
of the stars in these images using the IRAF task PHOT, which is part of 
the DAOPHOT package (Stetson 1987), and then we measured differential 
photometry for each MDEB with respect to several bright, constant 
check/comparison stars. Our science targets are somewhat redder than the 
typical comparison stars, so we tested for an airmass-dependent color term 
in the differential photometry, but found no evidence that one was needed 
at the level of our photometric precision. Finally, we compared the MDEB 
brightness from before and after each eclipse to test for systematic 
effects or secular changes in brightness due to starspots, but found 
little evidence for spot-driven brightness variations at a level of 
$\ga$1\%.

In Table 3, we list the total number of observations per filter for each 
eclipse window, the duration of the time series, and the estimated 
photometric precision. In Table 4, we list the 2833 brightness 
measurements from our P60 observations.

\section{Spectroscopic Observations}

\subsection{Long-Slit Spectroscopy}

\begin{deluxetable*}{lrrcrrrrrr}
\tabletypesize{\scriptsize}
\tablewidth{0pt}
\tablecaption{Low-Resolution Spectroscopic Observations}
\tablehead{\colhead{MG1-} & \colhead{Epoch} & \colhead{$t_{int}$} &
\colhead{SpT} & \colhead{EW(H$\alpha$)} & 
\colhead{$TiO_{7140}$} & \colhead{$TiO_{8465}$} & 
\colhead{$Na_{8189}$} & \colhead{$TiO5$} & 
\colhead{$CaH2$}
\\
\colhead{} & \colhead{(JD-2450000)} & \colhead{(s)} & \colhead{} & 
\colhead{(\AA)}
}
\startdata
78457&4083.72&300&M3.5&$-$6.5&1.69&1.12&0.85&0.46&0.51\\
116309&4083.75&300&M0.5&$-$2.9&1.15&1.02&0.93&0.79&0.72\\
506664&4084.03&420&M1.0&$-$2.0&1.30&1.03&0.91&0.61&0.59\\
646680&3906.66&300&M1.0&$-$3.3&1.23&1.00&0.92&0.56&0.55\\
1819499&3889.97&120&M0.5&$-$2.7&1.24&0.98&0.96&0.73&0.70\\
2056316&4083.66&300&M2.5&$-$2.3&1.46&1.06&0.87&0.52&0.49\\
\enddata
\end{deluxetable*}

As part of our selection process, we obtained moderate resolution optical 
spectra from our new MDEB systems. These spectra were intended to 
distinguish genuine low-mass eclipsing binary systems from various types 
of interlopers, but they can also be used to estimate the temperature, 
surface gravity, chromospheric activity, and (approximate) metallicity of 
the systems.

All of our spectra were obtained with the Double Spectrograph (Oke \& Gunn 
1982) on the Hale 5m telescope at Palomar Observatory. Spectra presented 
here were obtained with the red channel using a 316 l/mm grating and a 
2.0\arcsec\, slit, yielding a spectral resolution of $R\sim$1250 over a 
wavelength range of 6400-8800 angstroms. Wavelength calibration was 
achieved by observing a standard lamp after each science target, and flux 
normalization was achieved by periodic observation of spectrophotometric 
standard stars from the compilation of Massey et al. (1988). All spectra 
were dark-subtracted and flatfielded using standard IRAF routines, and 
then the stellar spectra were extracted using the IRAF routine APALL. 
Finally, wavelength calibration and continuum normalization were conducted 
using standard IRAF routines.

We list the epochs and exposure times in Table 5. Most of the spectra have 
$S/N > 50$, but the data for MG1-646680 are much noisier since it was 
observed in marginal conditions; we did not reobserve it because its 
quality was sufficient for spectral typing based on the broad TiO 
absorption bands.

\subsection{Echelle Spectroscopy}

In order to estimate the masses of our targets, we must measure the radial 
velocity curves of the individual component stars of each system. In a 
system with two 0.5 $M_{\sun}$ stars in a 1 day orbit, the orbital velociy 
of each star with respect to the other will be $\sim$200 km/s. We 
therefore must acquire high-dispersion spectra with velocity resolutions 
significantly lower than this value. Single-order spectrographs can 
achieve the required precision, but the preferred solution is to use 
echelle spectrographs that sample a much wider wavelength range by 
observing many orders of the spectra at once.

We obtained our high-dispersion spectra using the High-Resolution Echelle 
Spectrometer (HIRES) on the Keck-I 10m telescope. HIRES is a single-slit 
echelle spectrograph permanently mounted on the Nasmyth platform of Keck 
1. All observations were performed using the red channel of HIRES, and 
most span a wavelength range of 5300-9900 angstroms. All observations were 
obtained using the C2 or D1 deckers, which feature slit widths of 3 or 4 
pixels; the corresponding spectral resolutions are $R\sim$45000 or 
$R\sim$36000. We processed our HIRES data using the standard extraction 
pipeline MAKEE\footnote{http://spider.ipac.caltech.edu/staff/tab/makee}, 
which automatically extracts, flat-fields, and wavelength-calibrates 
spectra taken in most standard HIRES configurations.

In Table 6, we list the epochs and exposure times for all of our HIRES 
observations, as well as the $S/N$ for each spectrum at 6600 \AA. Due to 
the wide wavelength coverage in a typical echelle spectrum, radial 
velocities can typically be measured even when individual spectral lines 
are measured only at $S/N \la 3$. However, we obtained most of our 
observations at much higher $S/N$ in order to allow future measurement of 
individual line strengths, once suitable calibrations for properties like 
metallicity and surface gravity become available. As we show in Section 
5.3, this choice also allows us to achieve excellent precision in our 
final radial velocity measurements despite only having a small number of 
epochs. Two of our observing runs suffered from poor weather, so our phase 
coverage for some systems is not as even or dense as we would prefer. 
However, most systems with periods of $\la$7 days should tidally 
circularize within $\la$1 Gyr (e.g. Mathieu et al. 2004 and references 
therein), so we should only need 2 epochs (yielding 4 RVs) to constrain 
the three observational free parameters for each system: the RV curve 
amplitudes $K_A$ and $K_B$, plus the mean RV of the system. All additional 
epochs only serve to reduce the uncertainties as $\sqrt{N_{obs}}$ and to 
test for systematic noise due to spots, flares, and instrumental effects . 
They are also valuable in confirming that the orbits are truly circular.

During each sequence of observations, we also observed late-type stars 
from the list of RV standards compiled by Nidever et al. (2002). These 
standard stars were chosen to simultaneously serve as radial velocity 
standards and as fitting templates for our analysis of the MDEB spectra. 
We also observed a large number of FGK stars during twilight in order to 
further calibrate the RVs between nights. We list the K7-M4 standard stars 
that we used in Table 7. As was discussed by Nidever et al., even though 
these targets are RV stable at $\la$0.1 km/s over a timescale of a few 
years, the absolute RVs are likely to be systematically uncertain by 
$\sim$0.4 km/s at late spectral types.

We found that even when the instrument configuration is left unchanged, 
the velocity calibration can vary by $\sim$1 km/s over the course of a 
night. We have corrected this velocity drift in each observation by 
cross-correlating its telluric features (using the IDL \emph{c\_correlate} 
function) at 7600 \AA\, and 9300 \AA\, with the corresponding bands in 
five RV standards (GJ 450, GJ 908 (first epoch), GJ 408, HD 285968 (first 
epoch), and GJ 109) that appear to have zero velocity drift with respect 
to each other and to the average of all 20 remaining RV standards.

After applying the telluric RV correction, we cross-correlated all of our 
RV standards with each other in order to determine the intrinsic RV 
uncertainty for bright, slow-rotating M dwarfs. We found that the scatter 
for cross-correlations between pairs of spectra is $\sim$450 m/s, 
indicating that each spectrum has an intrinsic velocity uncertainty of 
$\sim$300 m/s. This measurement uncertainty is seen even between separate 
observations of the same targets, so it seems to be caused by 
astrophysical or instrumental effects, not uncertainties in the 
measurements by Nidever et al.. As we discuss in Section 5.3, we have 
calibrated each of our science observations with the $\sim$10 standard 
stars within $\pm$1 spectral subtype in order to reduce the calibrators' 
contribution to the error budget to $\sim$100 m/s; this contribution is 
small compared to the 300 m/s contribution from the science observations 
themselves, yielding total uncertainties of $\sim$350 m/s.

\begin{deluxetable*}{lcccrrrrr}
\tabletypesize{\scriptsize}
\tablewidth{0pt}
\tablecaption{High-Resolution Spectroscopic Observations}
\tablehead{\colhead{MG1-} & \colhead{Epoch} & \colhead{Phase} & 
\colhead{$t_{int}$} & \colhead{$S/N$} & 
\colhead{$v_{prim}$} & \colhead{$v_{sec}$} & 
\colhead{EW(H$\alpha$)$_{prim}$} & 
\colhead{EW(H$\alpha$)$_{sec}$}
\\
\colhead{} & \colhead{(HJD$-$2400000)} & \colhead{} & 
\colhead{(s)} & \colhead{(6600\AA)} & 
\colhead{(km s$^{-1}$)} & \colhead{(km s$^{-1}$)} & 
\colhead{(\AA)} & \colhead{(\AA)}
}
\startdata
78457&54484.89201&0.2453&900&6&$-$62.68&120.90&$-$2.99&$-$1.80\\
78457&54485.87340&0.8640&900&7&92.02&$-$46.23&$-$2.60&$-$1.70\\
78457&54485.93567&0.9033&900&6&75.65&$-$28.45&$-$2.77&$-$1.73\\
78457&54688.04238&0.3185&900&12&$-$55.10&111.45&$-$2.53&$-$1.91\\
78457&54688.09510&0.3520&900&13&$-$45.43&101.29&$-$2.13&$-$1.80\\
78457&54691.03629&0.2060&900&11&$-$59.29&116.88&$-$2.31&$-$1.89\\
78457&54691.09811&0.2450&900&10&$-$63.20&120.34&$-$2.47&$-$2.17\\
78457&54692.03966&0.8385&900&7&100.47&$-$55.44&$-$2.59&$-$1.85\\
78457&54692.10596&0.8803&900&6&86.32&$-$39.01&$-$2.46&$-$1.74\\
116309&54467.86931&0.0263&900&15&30.73&81.57&..&..\\
116309&54483.79741&0.2842&900&29&$-$54.34&173.11&$-$0.83&$-$1.28\\
116309&54483.83669&0.3317&900&32&$-$41.58&158.92&$-$0.88&$-$1.31\\
116309&54483.86564&0.3667&900&20&$-$25.58&143.52&$-$0.82&$-$1.32\\
116309&54485.92519&0.8568&900&26&142.45&$-$36.54&$-$0.91&$-$1.39\\
116309&54688.10473&0.2951&900&32&$-$53.75&172.23&$-$0.89&$-$1.24\\
116309&54691.13373&0.9568&900&23&86.78&20.17&..&..\\
116309&54692.12727&0.1579&900&17&$-$38.01&156.24&$-$0.92&$-$1.05\\
506664&54468.00623&0.7181&815&7&77.84&$-$111.26&$-$1.08&$-$0.91\\
506664&54468.01895&0.7266&900&12&77.95&$-$111.40&$-$0.94&$-$0.99\\
506664&54483.85152&0.9512&900&25&12.65&$-$40.87&..&..\\
506664&54485.96523&0.3163&900&26&$-$96.85&76.39&$-$1.15&$-$0.98\\
506664&54486.02658&0.3559&900&27&$-$84.73&63.31&$-$1.16&$-$0.93\\
506664&54486.07599&0.3878&900&24&$-$71.58&49.44&$-$1.11&$-$1.05\\
506664&54486.12058&0.4166&900&18&$-$57.87&34.67&$-$1.28&$-$1.19\\
646680&54466.99903&0.6339&1200&10&121.59&$-$11.73&$-$0.95&$-$0.63\\
646680&54467.16147&0.7331&1200&6&142.13&$-$35.09&$-$1.49&$-$0.81\\
646680&54485.95185&0.2071&900&13&$-$21.64&149.38&$-$1.58&$-$0.88\\
646680&54486.03798&0.2597&900&13&$-$24.24&151.82&$-$1.48&$-$0.77\\
646680&54486.10776&0.3023&900&14&$-$19.43&146.74&$-$1.18&$-$0.77\\
646680&54486.14463&0.3248&900&13&$-$14.51&141.36&$-$1.20&$-$0.79\\
1819499&54687.79186&0.1512&900&28&$-$124.05&90.63&$-$1.03&$-$0.99\\
1819499&54687.83039&0.2123&900&27&$-$141.40&108.04&$-$0.89&$-$0.97\\
1819499&54687.87370&0.2810&900&30&$-$139.30&107.20&$-$1.05&$-$0.99\\
1819499&54688.06275&0.5809&900&26&48.11&$-$87.91&$-$1.17&$-$1.12\\
1819499&54690.76360&0.8659&900&31&70.34&$-$111.34&$-$1.15&$-$1.40\\
1819499&54690.82930&0.9702&900&27&8.66&$-$38.70&..&..\\
1819499&54690.92684&0.1249&900&25&$-$110.93&77.40&$-$1.10&$-$1.19\\
1819499&54690.96773&0.1898&900&26&$-$137.02&104.04&$-$0.87&$-$0.96\\
1819499&54691.00505&0.2490&900&23&$-$143.14&110.50&$-$1.02&$-$1.01\\
1819499&54691.84209&0.5770&900&29&45.12&$-$86.06&$-$0.82&$-$1.07\\
1819499&54691.91316&0.6898&900&25&99.33&$-$142.97&$-$0.94&$-$1.08\\
1819499&54691.96015&0.7643&900&24&105.28&$-$147.06&$-$1.22&$-$1.26\\
2056316&54687.94116&0.1272&900&28&$-$58.51&63.69&$-$1.97&$-$0.90\\
2056316&54687.99439&0.1581&900&31&$-$67.79&74.56&$-$1.89&$-$0.85\\
2056316&54688.08689&0.2118&900&30&$-$77.42&86.45&$-$1.74&$-$0.83\\
2056316&54688.12519&0.2340&900&30&$-$78.95&88.52&$-$1.65&$-$0.75\\
2056316&54690.90102&0.8451&900&22&57.91&$-$79.25&$-$1.76&$-$0.94\\
2056316&54690.95446&0.8761&900&24&48.04&$-$67.49&$-$1.66&$-$0.92\\
2056316&54690.99224&0.8981&900&23&40.30&$-$57.81&..&..\\
2056316&54691.86303&0.4035&900&24&$-$45.41&47.66&$-$1.65&$-$0.88\\
\enddata
\end{deluxetable*}

\begin{deluxetable}{llrrr}
\tabletypesize{\scriptsize}
\tablewidth{0pt}
\tablecaption{RV/SpT Calibrators}
\tablehead{\colhead{Name} & \colhead{SpT} & \colhead{Epoch} &
\colhead{$v$} &  \colhead{$S/N$}
\\
\colhead{} & \colhead{} & \colhead{(HJD$-$2400000)} & \colhead{(km/s)} & 
\colhead{(6600\AA)}
}
\startdata
GJ 156&K7&54690&62.6&260\\
HD 28343&K7&54691&$-$35.1&85\\
HD 95650&M0&54485&$-$13.9&260\\
GJ 678.1A&M0&54691&$-$12.5&130\\
GJ 96&M0.5&54484&$-$37.9&300\\
GJ 908&M1&54466&$-$71.1&70\\
GJ 450&M1&54485&0.3&230\\
HD 165222&M1&54687&32.7&240\\
GJ 908&M1&54690&$-$71.1&270\\
HD 165222&M1&54691&32.7&140\\
HD 36395&M1.5&54485&8.7&340\\
HD 36395&M1.5&54687&8.7&250\\
HD 216899&M1.5&54691&$-$27.3&160\\
GJ 2066&M2&54467&62.2&55\\
HD 285968&M2.5&54467&26.2&110\\
GJ 408&M2.5&54485&3.2&220\\
HD 180617&M2.5&54687&35.9&250\\
HD 180617&M2.5&54691&35.9&140\\
HD 285968&M2.5&54691&26.2&70\\
HD 173739&M3&54690&$-$0.8&160\\
GJ 109&M3.5&54484&30.6&190\\
HD 173740&M3.5&54690&1.2&100\\
GJ 628&M3.5&54691&$-$21.2&100\\
GJ 447&M4&54485&$-$31.1&130\\
GJ 699&M4&54691&$-$110.5&130\\
 \enddata
 \tablecomments{Velocities were adopted from the list of standard stars 
reported by Nidever et al. (2002), who found that they have internal 
dispersion of $\la$0.1 km/s and systematic uncertainties of $\la$0.4 km/s. 
Spectral types were adopted from the PMSU surveys (e.g. Reid et al. 
1995).}
 \end{deluxetable}

\section{Analysis and Results}

Our data analysis can be divided into several major stages, which we 
describe in the following subsections. First, we analyzed the 
moderate-resolution spectra of each system in order to characterize their 
atmospheric properties and confirm that they should have component masses 
$M\la$0.6 $M_{\sun}$. Next, we combined all of the time-series photometry 
of our systems in order to measure their orbital periods and eclipse 
timing. After this, we analyzed the high-dispersion spectra of each system 
in order to measure their radial velocity curves and component masses 
(modulo inclination). Finally, we analyzed the multicolor eclipse light 
curves of our systems in order to measure the components' masses, radii, 
and temperatures.

\subsection{Spectral Types and Emission Line Strengths}

 \begin{figure}
 \plotone{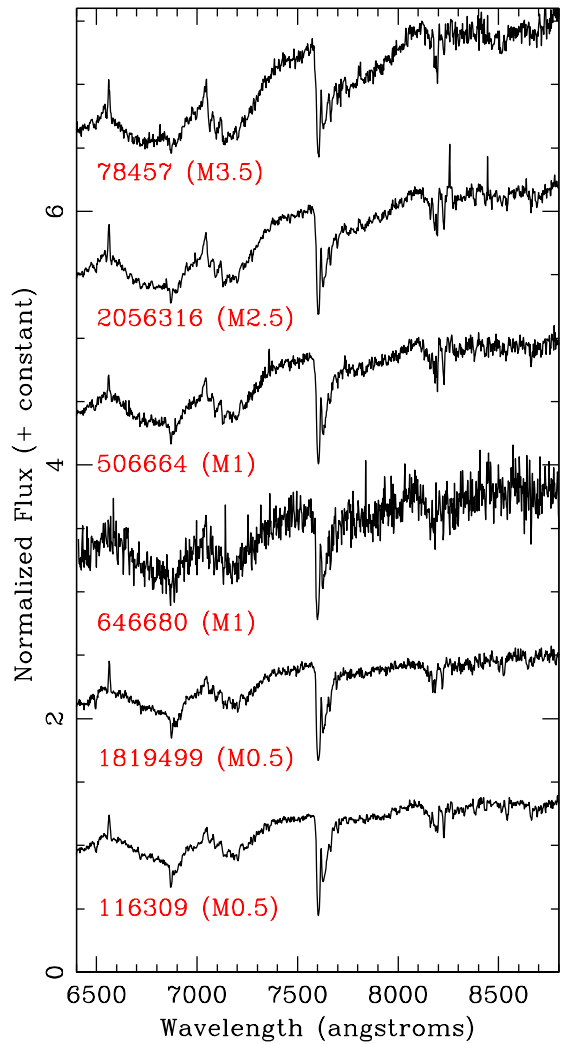}
 \caption{Intermediate-resolution spectra for our six new MDEB systems. 
The TiO band at 7100 angstroms is clearly present in all spectra, 
suggesting spectral types of M0 or later. Most also show some evidence of 
Halpha emission, which indicates either youth or chromospheric activity. 
None possess the shallow Na-8189 doublet that is characteristic of young 
systems and most are not near regions of ongoing star formation, so the 
H$\alpha$ emission seems to result from as the high activity common to 
close binary systems. The spectrum of MG1-646680 is noisier, despite its 
longer integration time, because it was observed in marginal conditions. 
(A color version of this figure is available in the online journal.)}
 \end{figure}

The moderate-resolution spectra yield spectral types that allow us to 
determine temperatures in a way that is independent of our broadband SED 
fitting (Section 2). Furthermore, the depths of alkaline absorption lines 
can demonstrate low surface gravity indicative of youth (Slesnick et al. 
2006a), and the relative strengths of molecular bands (e.g. metal hydride 
versus metal oxide) distinguish metal-poor subdwarfs from 
solar-metallicity dwarfs (Woolf \& Wallerstein 2006).

In Figure 1, we plot the flux-normalized spectra for each of the new 
systems. We estimated spectral types via qualitative comparison of each 
spectrum to a range of standard stars from the work of Slesnick et al. 
(2006a, 2006b), who used the same instrument configuration. We confirmed 
these estimates by calculating the spectral indices TiO-7140 and TiO-8465, 
which measure the depth of key temperature-sensitive features (Slesnick et 
al. 2006a). We find that these indices are consistent with our assigned 
spectral types, but this only provides a strong constraint for sources 
with types of $\ga$M2 stars since both indices saturate for types earlier 
than $\sim$M1.

Our SED-fit spectral types are consistent with the spectroscopic spectral 
types, but typically more precise, so we used the SED-fit measurements to 
determine the effective temperatures for each component of each binary. 
For each system, we first used our composite spectral type (which 
represents the average of both components) to estimate a composite 
temperature from the temperature scale we reported in Kraus \& Hillenbrand 
(2007). We then specifically estimated the component temperatures using 
the temperature ratio we inferred in Section 5.4, assuming that the 
composite temperature represented an average of the two component 
temperatures that was weighted by their respective contributions to the 
total system luminosity.

We also measured the equivalent width of H$\alpha$ emission, a measure of 
chromospheric activity. As we discuss in Section 5.3, all of our targets 
show $H\alpha$ emission, matching the ubiquity seen for most short-period 
spectroscopic binaries. However, the emission was not always strong enough 
to be measured in these lower-resolution spectra, especially for targets 
which were observed at low $S/N$. Finally, we measured the 
gravity-sensitive spectral index Na$_{8189}$ (Slesnick et al. 2006a) and 
the metallicity-sensitive spectral indices TiO5 and CaH2 (Woolf \& 
Wallerstein 2006). In all cases, these spectral indices are consistent 
with dwarf gravity and near-solar metallicity.

We summarize all of these measurements in Table 5.

\subsection{Eclipse Timing}

 \begin{figure*}
 \plotone{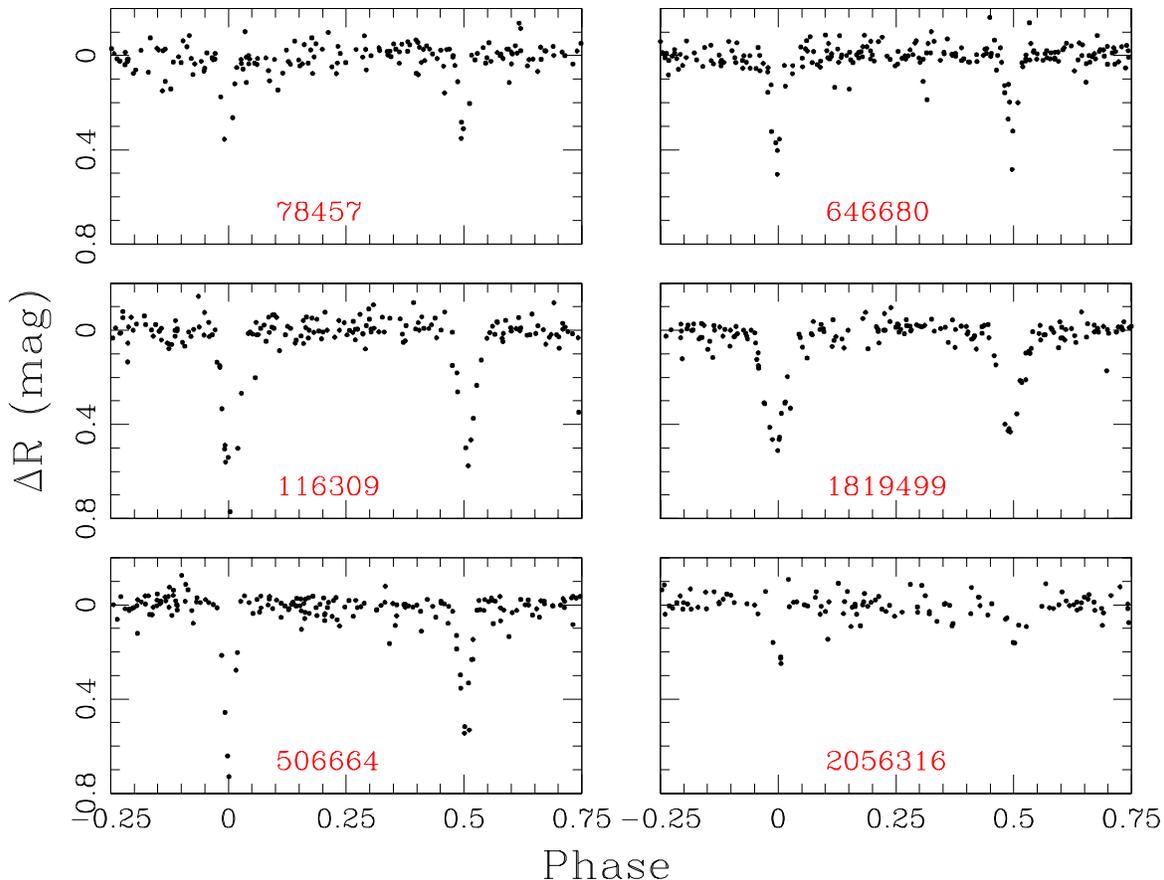}
 \caption{Phased discovery light curves for our systems, as measured with 
the MOTESS telescopes. All of the systems showed evidence of eclipses in 
the unphased light curves, but the discovery light curves did not have a 
sufficient number of points to determine an unambiguous period. (A color 
version of this figure is available in the online journal.)}
 \end{figure*}

 \begin{figure*}
 \plotone{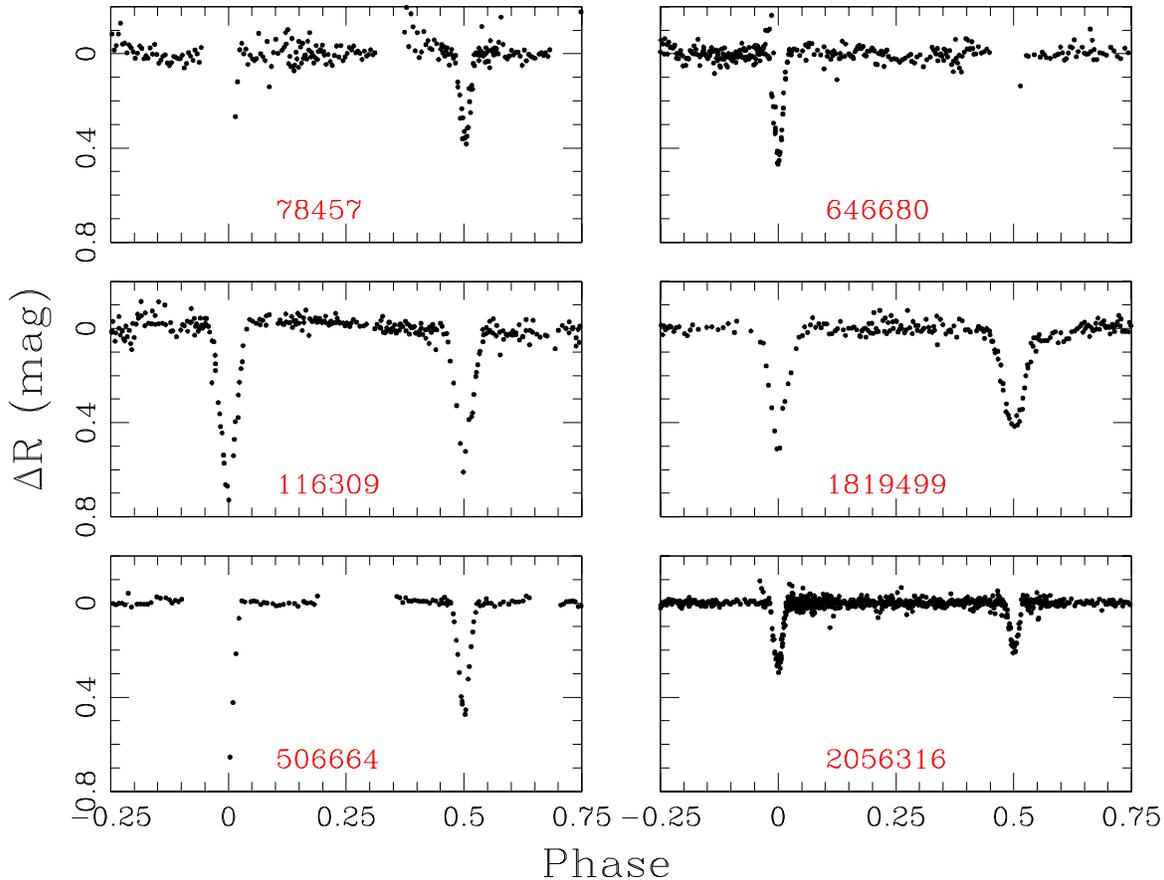}
 \caption{Phased followup light curves for our systems as measured with 
several small telescopes. These observations typically spanned several 
whole nights in a row, and the detection of several consecutive eclipses 
allowed us to determine each system's actual period; we then combined this 
data with our MG1 observations to determine the system periods and eclipse 
epochs needed for additional followup observations. (A color version of 
this figure is available in the online journal.)}
 \end{figure*}

 \begin{figure*}
 \plotone{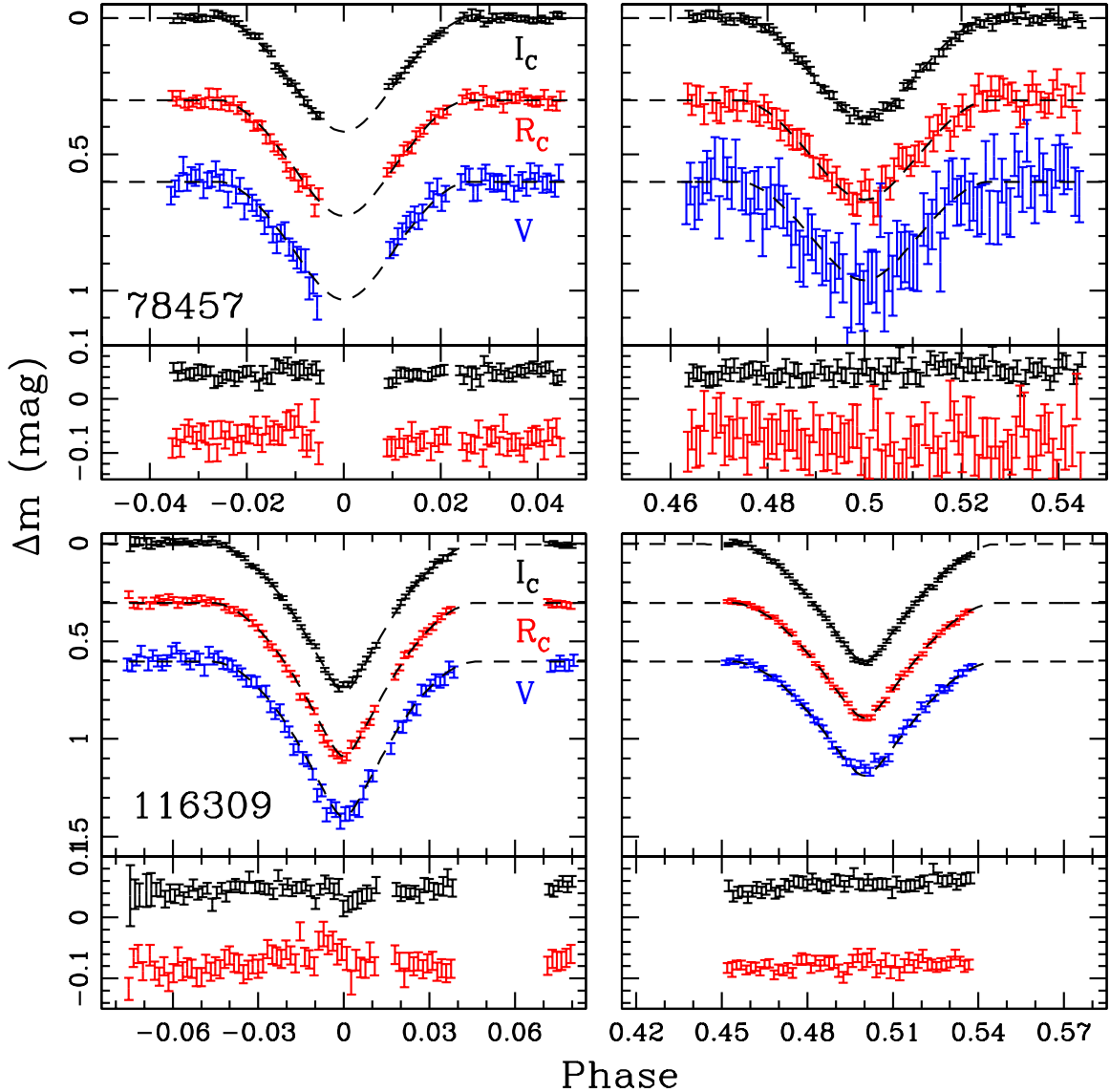}
 \caption{Multicolor eclipse light curves for two of our newly-discovered 
M dwarf eclipsing binary systems: MG1-78457 and MG1-116309. For each 
system, we show the light curves in $I_C$ (black), $R_C$ (red), and $V$ 
(blue), as well as the predicted light curves from our best-fit radius 
models (dashed lines). The $R_C$ and $V$ observations were offset to avoid 
overlap. Below each plot, we show the residuals in $I_C$ and $R_C$ with an 
expanded scale in order to demonstrate the typical scatter; we do not show 
the residuals for V in order to avoid crowding the plot and because the 
typical scatter can be discerned adequately without an expanded scale. As 
we discuss in the text, a flare was seen in the middle of the secondary 
eclipse for MG1-78457; we have omitted those data points in fitting our 
light curve models. (A color version of this figure is available in the 
online journal.)}
 \end{figure*}

 \begin{figure*}
 \plotone{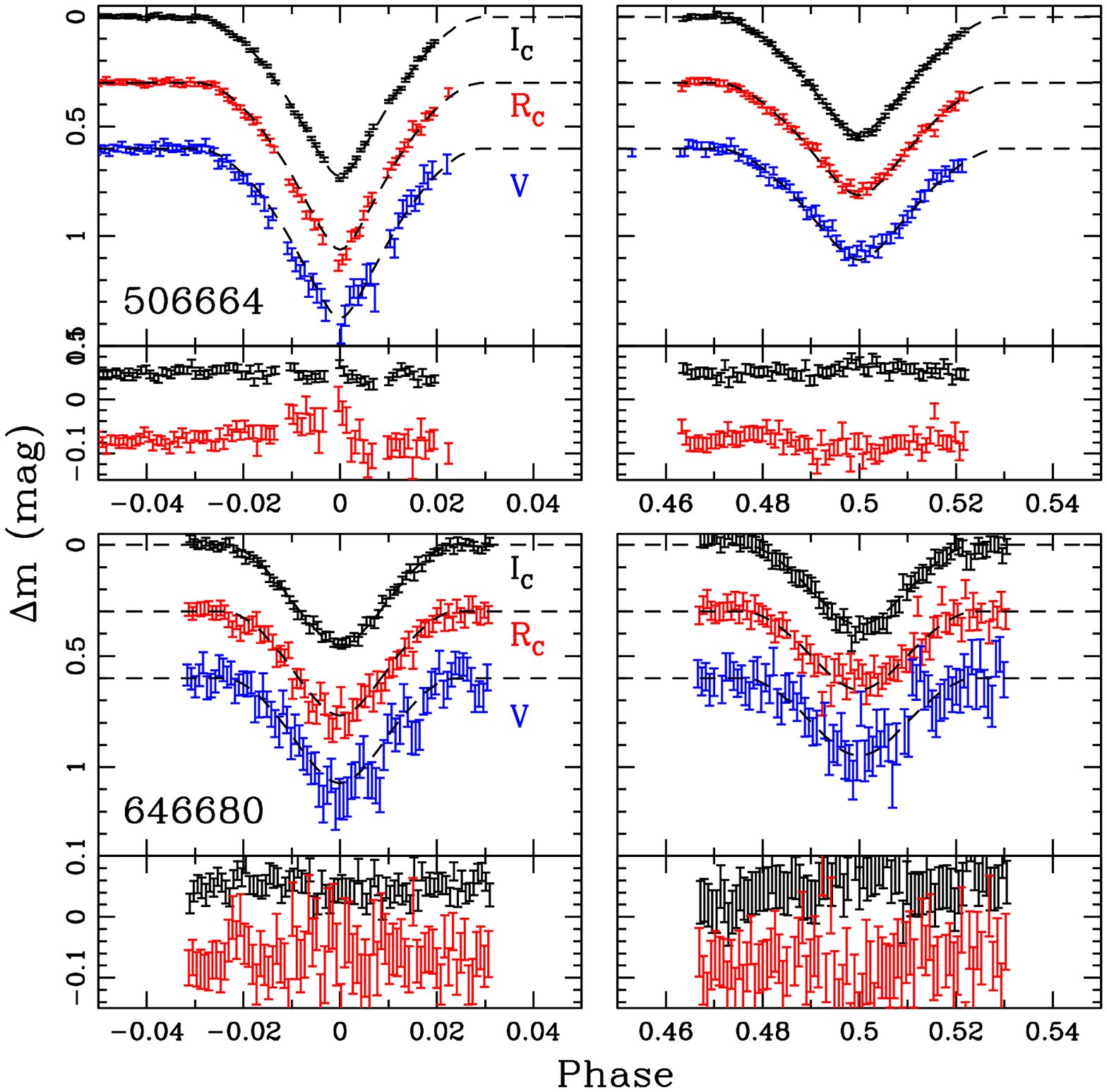}
 \caption{Multicolor eclipse light curves for two more newly-discovered M 
dwarf eclipsing binary systems: MG1-506664 and MG1-646680. Figure layout 
and labels are the same as for Figure 4. (A color version of this figure 
is available in the online journal.)}
 \end{figure*}

 \begin{figure*}
 \plotone{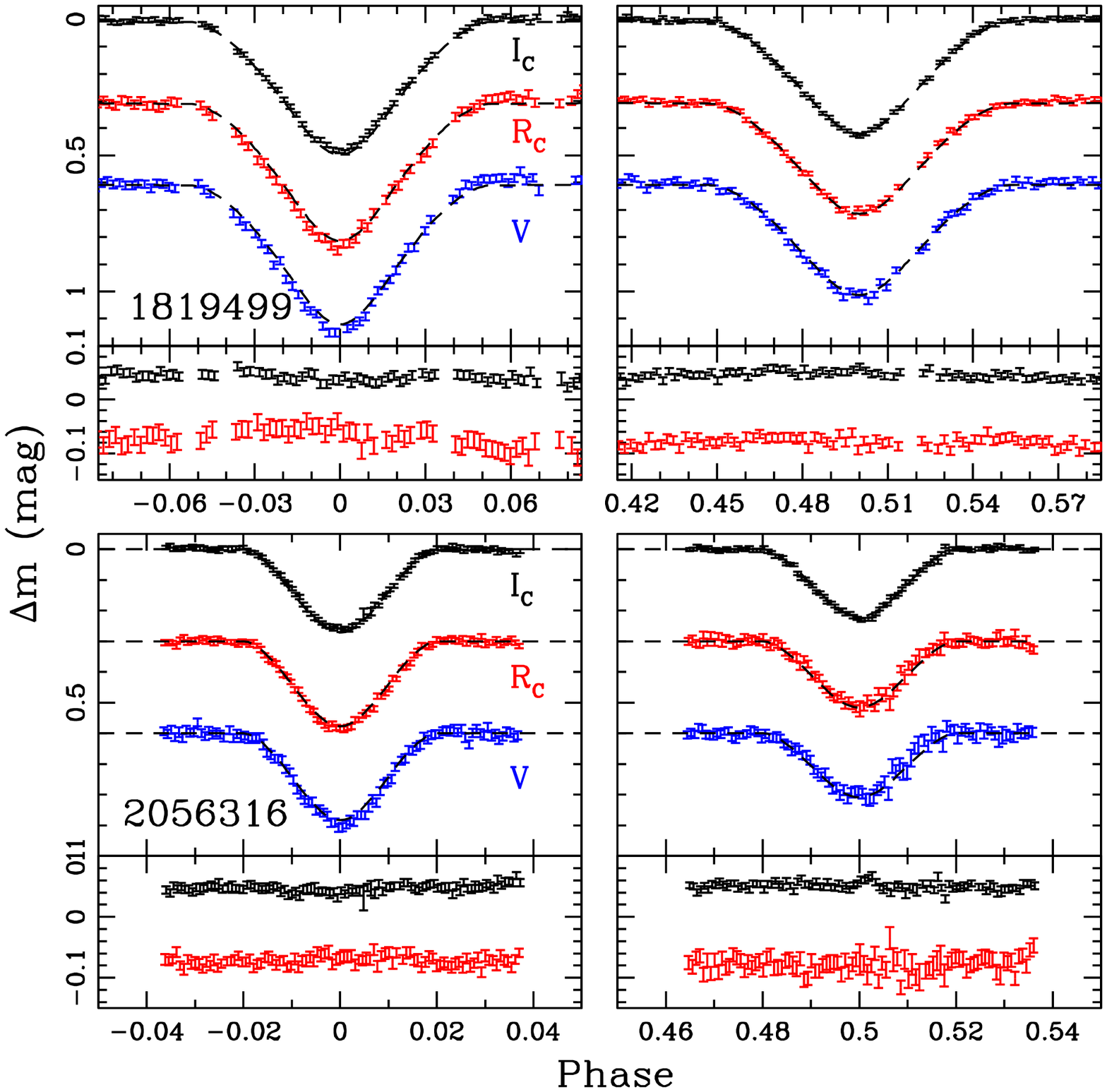}
 \caption{Multicolor eclipse light curves for two more newly-discovered M 
dwarf eclipsing binary systems: MG1-1819499 and MG1-2056316. Figure layout 
and labels are the same as for Figure 4.(A color version of this figure is 
available in the online journal.)}
 \end{figure*}

Our photometric observations span an interval of $\sim$7 years, so 
simultaneous analysis of all data should yield very precise measurements 
of the orbital periods ($P$). The fast cadence and high precision of the 
P60 observations should also allow us to estimate the time of eclipse 
($t_0$) very precisely. However, our data is heterogeneous and represents 
several different telescopes and filter systems, so our analysis proceeds 
through several steps in order to achieve this result. Our final results 
show that the measurement of $t_0$ is set entirely by our data from the 
P60, while the measurement of $P$ is set by that value of $t_0$ and by the 
eclipse epochs in our MG1 and small-telescope followup data.

For each system, we first used the period analysis package 
PERANSO\footnote{http://www.peranso.com/} to analyze the light curve from 
our small telescope followup observations, measuring the orbital period 
with a precision of $\la$0.01d and establishing a provisional primary 
eclipse epoch with a precision of $\sim$2--5 minutes. This step was 
critical because aliasing and sparse time coverage made it impossible to 
determine unambiguous periods from MG1 data alone, and we required those 
periods to plan all subsequent followup observations. After determining 
provisional values, we then exploited the much longer time baseline of the 
MG1 light curves (both alone and in combination with our followup data) to 
refine our estimate of each system's period with a precision of $\la 
10^{-5}$ d. We also used these updated values to plan optimal followup 
observations with the P60. Finally, we used the P60 eclipse light curves 
to establish a final value of $t_0$, then reanalyzed the MG1 and 
small-telescope data with this fixed value of $t_0$ in order to measure a 
final value for the orbital period (with a precision of $\sim 10^{-6}$ d).

We summarize our final periods and eclipse epochs for each system in Table 
8, and in Figures 2--6 we show the final phased light curves for MG1 
observations (Figure 2), our small telescope followup (Figure 3), and the 
P60 multicolor observations (Figure 4, Figure 5, Figure 6). We found that 
the typical uncertainty in the P60 eclipse timing (as determined from the 
dispersion between filters and between the primary and secondary eclipses) 
is $\sim$10--20 sec ($\sigma_{t0} /P \sim 10^{-4}$), while the 
uncertainties in the periods are $\sim$0.05--0.3 seconds ($\sigma_P /P 
\sim 10^{-6}$--$10^{-7}$). Even though the eclipse timing from the 
earliest observations has relatively modest accuracy ($\sim$5--10 
minutes), the corresponding period is still very precise since the 
uncertainty in the period declines with the inverse of the number of 
periods that our dataset spans (typically $\ga$2000-5000).

\subsection{Radial Velocities}

 \begin{figure*}
 \plotone{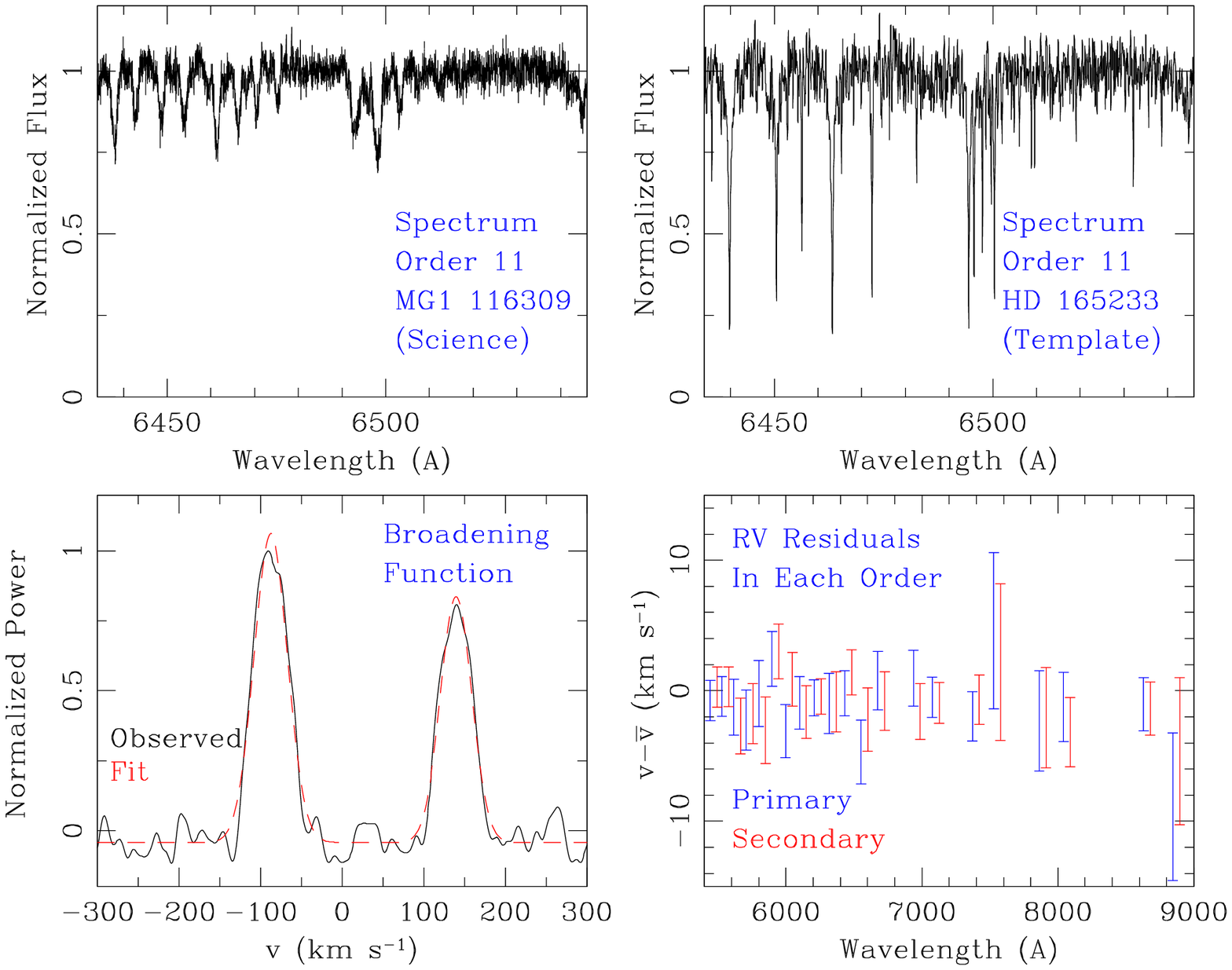}
 \caption{Analysis and results for one HIRES spectrum of MG1-116309. Upper 
Left: One order of the echelle spectrum for MG1-116309, as observed on JD 
2454688.1. Upper Right: The same echelle order for an RV standard star, HD 
165233, which was observed earlier that night. Lower Left: The broadening 
function (black solid line) which, when convolved with the RV standard 
spectrum, yields the best fit to the science spectrum. The two peaks are 
well-fit with a pair of Gaussian functions (red dashed line) where the 
mean of each Gaussian corresponds to the RV difference between that 
component and the RV standard. Lower Right: The residuals as a function of 
wavelength for the primary (blue) and secondary (red) RV measurements in 
each order, as measured with respect to the weighted mean of all 
measurements. Since the residuals often overlap, we offset the secondary 
point by 0.01 phase. (A color version of this figure is available in the 
online journal.)}
 \end{figure*}

 \begin{figure*}
 \plotone{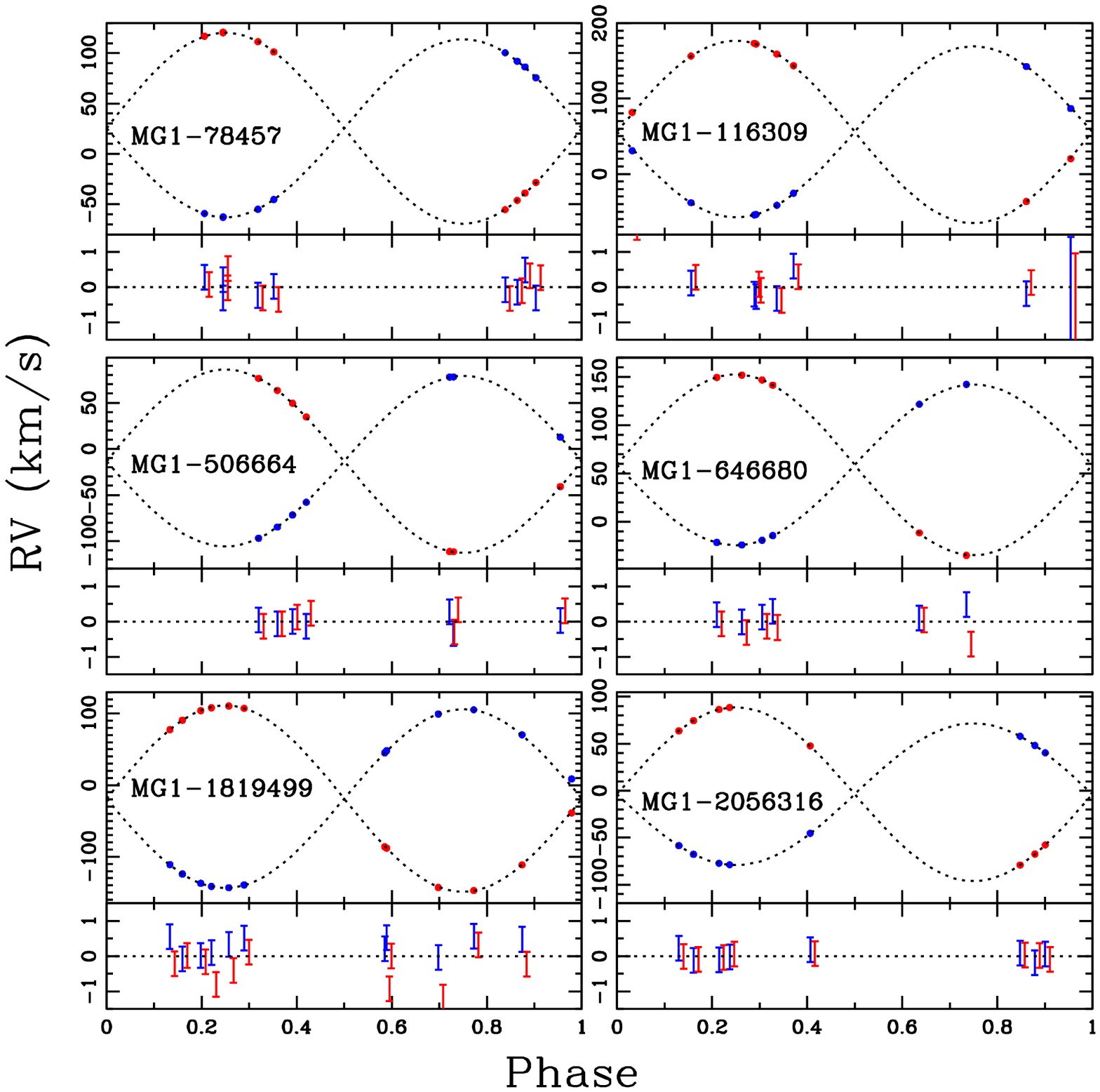}
 \caption{Radial velocity curves for our six new M dwarf eclipsing 
binaries. For each system, we plot the observed RVs of the primary (blue) 
and secondary (red), as well as the best-fit sinusoidal RV curves for that 
system. Underneath each plot, we also show the residuals around the 
best-fit model. (A color version of this figure is available in the online 
journal.)}
 \end{figure*}

There are several methods commonly used to analyze high-dispersion spectra 
and measure radial velocities, including the cross correlation, 
two-dimensional correlation (TODCOR; Mazeh \& Zucker 1994), and broadening 
function deconvolution (BF; Rucinski et al. 1999). We have chosen to 
analyze our data using BF; as Rucinski et al. described, BF is less 
susceptible to effects like ``peak pulling'' than correlation techniques, 
especially for targets like ours where the cross-correlation peaks are 
broadened by rotation and might partially overlap. S. Rucinski distributes 
an IDL pipeline that is designed to conduct BF for any input 
spectrum\footnote{http://www.astro.utoronto.ca/$\sim$rucinski/}, and we 
adopted this pipeline as written.

For each order of each spectrum, we used the BF pipeline to calculate the 
broadening function with respect to a bright RV standard star of similar 
spectral type. We then fit the two peaks of the broadening function with a 
pair of Gaussian functions in order to measure the component RVs for that 
order, and measured the average component RVs for that epoch by 
calculating a weighted mean of all orders. We estimated those weights by 
iteratively calculating the weighted mean RVs for all observations, 
measuring the standard deviation of each order's residuals around those 
averages, and using the standard deviations to update the weight assigned 
to each order. We started by assigning all orders an equal weight, and 
then iterated until the weights and the average component RVs converged. 
Finally, we repeated this process for all RV standards within $\pm$1 
spectral subclass of the science target (Table 5), and averaged the 
resulting RVs at each epoch in order to minimize systematic uncertainties 
in the RV calibration.  A typical cross-correlation to a single standard 
incurs velocity errors of $\sim$300 m/s from the science target and 
$\sim$300 m/s from the RV standard, but by using $\sim$10 RV standards, we 
can reduce the second contribution to a negligible $\sim$100 m/s. We find 
that the scatter of our measurements around the RV orbit fits is $\sim$300 
m/s; including degrees of freedom that contribute to the fit, then the 
inferred uncertainties are $\sim$350 m/s, which is consistent with our 
estimate.

In Figure 7, we demonstrate the steps of this process by plotting one 
echelle order for an observation of MG1-116309, the corresponding echelle 
order of one RV template star, the broadening function that relates the 
standard star spectrum to the spectrum of MG1-116309, and the RV residuals 
as a function of wavelength for each order of that spectrum (as determined 
from all epochs). Despite the significant rotational broadening seen for 
both components of MG1-116309, the order shown (and many other orders) 
yield measurements with an individual precision of $\sim$1--2 km s$^{-1}$. 
However, many of the orders suffer from severely degraded precision; this 
is typically a sign that those orders have fewer deep spectral lines than 
the orders that offer better precision. We list the component RVs (as 
averaged over all orders, and then over all RV standards) for each epoch 
in Table 6.

All of our new binary systems have orbital periods that are significantly 
shorter than the canonical limit for tidal circularization (7--10 days; 
Zahn 1977), so they should have zero eccentricity. This is consistent with 
the results of our photometric monitoring, which show that all secondary 
eclipses are displaced by half of the period from the primary eclipse. 
Given the assumption of circularity, each system's RV curve can be 
described by a pair of sinusoidal functions with four free parameters: the 
system period ($P$), the mean system RV ($\gamma$), and the amplitude of 
each component's RV curve ($K_A$ and $K_B$). In each case, we hereafter 
subscript these quantities with ``p'' to denote the primary star and ``s'' 
to denote the secondary star. Since we know the orbital periods from the 
systems' light curves (Section 5.2), we can further simplify our RV 
analysis by adopting the previously-measured value of $P$ and only fitting 
for the other three parameters. 

For each of our targets, we measured the best-fit values and uncertainties 
for $\gamma$, $K_A$, and $K_B$ via a $\chi^2$ minimization of the 
double-sinusoid model:

 \begin{eqnarray}
 &v_p& = \gamma - K_A \times \sin{(2 \pi \theta)} \\
 &v_s& = \gamma + K_B \times \sin{(2 \pi \theta)}
 \end{eqnarray}

In Figure 8, we show our phased RV observations and the corresponding 
best-fit models. The models and data show excellent agreement, with a 
typical dispersion of $\sim$300 m/s for all RV measurements where the 
component spectra were well-resolved (i.e. occurring $>$0.05 phase from an 
eclipse). There are only a small number of measurements that occurred 
while the system was near an eclipse, and most are also affected by the 
Rossiter-McLaughlin effect (Rossiter 1924; McLaughlin 1924). It appears 
that the uncertainties climb to $\sim$2.5 km s$^{-1}$ in this regime where 
the two components' spectra are not clearly resolved. Finally, as we noted 
above, the total system mass depends on $K_A-K_B$, which in turn can be 
fit from the velocity difference at each epoch, $(K_A+K_B) = (v_p-v_s) 
\times \sin{(2 \pi \theta)}$. Since the measurements are treated as a 
difference of two simultaneous measurements ($v_p-v_s$), then any 
epoch-to-epoch systematic uncertainties would cancel and allow for a more 
precise mass measurement. However, we found that the resulting masses have 
similar uncertainties as if we fit for the component masses and combine 
them. This indicates that the stochastic RV uncertainties are of similar 
order as the systematic RV uncertainties, and thus that further reduction 
of the systematic effects is unlikely to improve our measurements.

We summarize our measurements of the observed model parameters ($K_A$, 
$K_B$, and $\gamma$) and the corresponding physical system parameters 
(component masses $M_A \sin^3 (i)$ and $M_B \sin^3 (i)$, orbital semimajor 
axis $a \sin(i)$, and component mass ratio $q$) in Table 8. In all cases, 
star A is the more massive (primary) star that is eclipsed (with a deeper 
amplitude) at phase 0.0 and has a velocity amplitude of $K_A$.

\subsection{Eclipse Fitting}

We performed our eclipse-fitting analysis using the 2007 release of the 
venerable Wilson-Devinney code (WD; Wilson \& Devinney 1971), which 
produces synthetic lightcurves in a specified set of bandpasses based on a 
user-defined set of input system parameters. The WD code is packaged with 
a fitting routine that computes a converging best-fit solution based on 
differential corrections to an initial estimated solution. However, we 
found that this fitting routine did not converge well for our data, most 
likely because we have few observations outside of the eclipses. We 
instead used the WD code to produce synthetic lightcurves, then computed 
the $\chi ^2$ goodness-of-fit as compared to our data. While computing our 
fits, we fixed many of the system parameters ($P$, $q$. $a \sin(i)$, and 
$T_{eff,A}$) to the values computed in the previous sections, and 
continued to assume that the orbits have been tidally circularized. We 
adopted a square root limb darkening law as prescribed by van Hamme 
(1993), including the appropriate temperature-dependent exponents for each 
component. The only remaining parameters, which we solved for using the WD 
code, are the orbital inclination $i$, the component radii $r_A$ and 
$r_B$, and the difference in component temperatures $\Delta T$.

We adopted an iterative procedure for finding the best fit. We began by 
fixing $r_A$ and $r_B$ to a series of initial estimates corresponding to 
fractional values above and below predictions from theoretical models, 
with initial values of 95\%, 100\%, 105\% and 110\% of the theoretical 
radius for that given mass. We then performed a grid search over all 
values of $i$ and $\Delta T$ to find the best fit for those radii. After 
we had found the best possible solutions for each set of assumed radii, we 
then relaxed the radius constraints and allowed all four parameters to 
vary, computing differential corrections in order to converge into the 
minimum of the $\chi^2$ space. Finally, after we had found a minimum, we 
performed a grid search of all four parameters around that position in 
order to confirm that it was a global minimum rather than a local minimum. 
We accepted a solution only after all four of our initial radius estimates 
ultimately converged into that minimum, which occurred promptly in all 
cases and yielded reduced $\chi^2$ values of $\chi_{\nu}^2 =$ 0.9--1.9. We 
computed the uncertainties in our best-fit values of $R_A$, $R_B$, $i$, 
and $\Delta T$ from the shape of the $\chi^2$ surface around that minimum. 
We summarize the results of this fitting process in Table 8, and, we plot 
the best-fit models for comparison to each observed light curve in Figure 
4, Figure 5, and Figure 6.

In the majority of cases, we found that the formal uncertainty in the 
best-fit solution was extremely small. The only exceptions were for 
MG1-78457 and MG1-646680, where we found radius uncertainties of 
$\sim$2\%. This is a natural consequence of either losing part of an 
eclipse (as for MG1-78457, due to a flare) and/or using low-quality data 
(as for MG1-646680, where both eclipses were observed in marginal 
conditions). In general, the very small uncertainties in our results are 
not surprising given the volume and precision of our data. Each system had 
$\sim$300--600 photometric observations spread between three filters and 
two eclipse windows, and the uncertainties were $\la$0.01 mag in $I_C$ and 
$\la$0.05 mag even in V. However, these small stochastic uncertainties 
suggest that our true uncertainties could be driven by systematic effects 
rather than stochastic errors. Several different software packages are 
commonly used to fit light curves (e.g. Wilson \& Devinney 1971; Popper \& 
Etzel 1981; Southworth et al. 2004; Prsa \& Zwitter 2005), so it seems 
possible that the choice of algorithm could be significant. However, we 
used our procedures to re-fit the $R_C$ and $I_C$ light curves reported 
for GU Boo by Lopez-Morales \& Ribas (2005), and given the same spot 
model, we found the same parameters to within $\sim$1\%. As long as other 
authors have conducted similar tests, then we expect that this systematic 
uncertainty will be of similar order. We have also reported all of our 
photometry and radial velocities in order to allow future calibration of 
our results against other algorithms.

Otherwise, the most likely source of systematic uncertainty may be in the 
likely presence of starspots that can introduce extra variability. 
Previous observations of low-mass eclipsing systems like GU Boo have found 
that the inferred radii can change by $\sim$1--2\% depending on the 
details of spot modeling (e.g. Lopez-Morales \& Ribas 2005 versus Morales 
et al. 2009); observations of other systems also find spots to be 
significant at this level (e.g. Irwin et al. 2009; Windmiller et al. 
2010). A model for the spot distribution can be inferred from a 
well-sampled light curve that covers most of an eclipsing system's orbital 
period; since these stars are tidally locked, then the spots modulate the 
overall brightness of the system on the same period as the orbital period, 
and the phase of these modulations sets the longitudinal distribution of 
spots. However, we observed our systems only around the time of eclipse, 
so we do not have sufficient information to construct a spot model. We 
knew before conducting these observations that uncertainties in the spot 
configuration would likely determine the ultimate accuracy of our 
observations, but since we had limited resources and many targets to 
observe, we decided to accept this limitation for the purposes of 
preliminary characterization. Much of the effect should be removed when we 
renormalize our eclipse light curves using the out-of-eclipse brightness, 
but we still must determine the level at which remaining variability can 
influence our results.

To this end, we have conducted a set of ``artificial spot'' experiments. 
For each system in our sample, we have constructed an artificial 
lightcurve matching the measured system properties and sampled at the same 
epochs as our observations. We then introduced various spot configurations 
into these artificial systems and attempted to refit the light curves with 
spotless models, thereby measuring the effect on the best-fit system 
properties. It is still unclear what sets the configuration of spots, so 
instead of adopting random models, we adopt the spot latitudes, 
longitudes, sizes, and temperature ratios inferred for various epochs for 
CM Dra (M4; Morales et al. 2009) and GU Boo (M0; Windmiller et al. 2010).

We found that uncorrected spots typically led to a dispersion in the 
inferred system parameters of $\sim$0.002-0.023 $R_{\sun}$ for $R$, 
$\sim$0.05-0.20$^o$ for $i$, and $\sim$8--14 K for $\Delta T$. In general, 
systems are affected more significantly if the eclipses are quick and 
shallow (as for MG1-1819499). Larger uncertainties are also incurred if 
there are fewer observations taken before and after the eclipses (as for 
MG1-116309) since there is a higher uncertainty in renormalizing the 
constant brightness offset from the spot. In cases like MG1-2056316 or 
MG1-646680, where both eclipses are bracketed by well-sampled 
observations, the effect of the spots on the radii is minimal. The effect 
on $\Delta T$ is more significant, and for good reason; if a nontrivial 
fraction of a star's surface is covered by a spot or plage, then its 
average temperature is indeed higher or lower than in the unspotted case.

We list the apparent spot-related uncertainties for each system's derived 
properties in Table 8. However, we suspect that these systematic 
uncertainties might be overestimated. Our systems have rotational periods 
that are a factor of $\sim$2--5 slower than most of the previously studied 
MDEBs, so they might be less active and show fewer and smaller spots. This 
is corroborated by our light curves, as most show consistent brightnesses 
even on very long timescales (e.g. Section 3.1; Figure 2). As we discuss 
in Section 6, the inferred radii for eight components in four of our 
systems with similar periods (1.5--1.8 days) and masses (0.35-0.60 
$M_{\sun}$) have model-normalized radii that are typically consistent to 
within $\pm$2\%, which also strongly argues that systematic effects are 
not significant. However, the components of short-period binary systems do 
not show such consistency. These systems might be affected by spots, 
though the similar inconsistency for stars with well-determined spot 
models (like GU Boo; Morales et al. 2009) suggests that there could be a 
genuine dispersion in stellar radii.

Finally, we note that light curve fits generally provide a much stronger 
constraint on the sum of the radii (which corresponds to the eclipse 
duration) than on the individual component radii (which only affect the 
detailed shape of each eclipse). As a result, light curve fits face a 
degeneracy between the component radii unless the photometry is very 
precise. We characterized this requirement by using the WD algorithm to 
simulate systems with the same radius sum ($R_A+R_B$) and different 
individual radii, and found that a change of 1\% in the component radii 
corresponded to a difference of $\sim$2 mmag for each point in the eclipse 
light curve. Our multicolor eclipse lightcurves typically contain 
$\sim$100-200 $R_CI_C$ points with precisions of $\sim$10 mmag, so the 
overall fit should allow us to distinguish radii at this precision. Our 
results could be systematically incorrect if our light curves include red 
noise (e.g. Pont et al. 2006), but we have verified that bright constant 
stars in our fields are typically precise to $\sim$3--4 mmag, so this 
suggests that any systematic uncertainty in the radii should have a 
magnitude of $\la$2\%, and hence be comparable to that from unmodeled 
spots.

Previous studies have also broken this degeneracy by invoking the observed 
flux ratio from spectral fitting of the high-resolution spectra used for 
measuring RV curves (e.g. Stassun et al. 2008; Irwin et al. 2009). If the 
flux ratio is known at a given wavelength, then the appropriate bolometric 
corrections can be applied in order to infer the system luminosity ratio; 
this value can then be combined with the observed temperature ratio in 
order to infer the ratio of the component radii, and hence the individual 
radii. However, this technique faces a fundamental systematic limit from 
our imperfect knowledge of bolometric corrections (which we estimate at 
$\sim$5\% for early M dwarfs; e.g. Leggett et al. 1996), as well as a 
stochastic limit from the low precision of flux ratios which are derived 
from rotationally broadened spectra (which we observe to be $\sim$5\% for 
our HIRES data). Despite these limits, we tested this technique for our 
targets by measuring the ratio of the component fluxes around the CaII 
infrared triplet, which falls near the blackbody peak and is only very 
weakly temperature-dependent on scales of $\sim$50--100 K (e.g. Cenarro et 
al. 2002), and then translating the flux ratio to a luminosity ratio by 
using interpolations of the broadband bolometric corrections we tabulated 
in Kraus \& Hillenbrand (2007), and finally computing radius ratios and 
component radii by invoking the temperatures we measured in our light 
curve fits. These spectroscopically measured component radii are 
consistent with the photometrically measured component radii to within 
$\sim$3--4\%, which matches the scatter expected from observational and 
systematic errors in this process. However, our light curve fits typically 
yield errors which supercede this level of agreement, so we will not use 
the spectroscopic constraints on the component radii in our subsequent 
analysis.

\section{The Mass-Radius(-Period?) Relation for Low-Mass Eclipsing 
Binaries}

The past decade has seen a revolution in the calibration of fundamental 
stellar properties. Binary orbit monitoring has yielded many new dynamical 
mass measurements (Delfosse et al. 2000; Balega et al. 2005; Martinache et 
al. 2007; Dupuy et al. 2009), precise parallaxes have led to the 
measurement of highly precise luminosities (Deacon et al. 2005; Henry et 
al. 2006; Lepine et al. 2009), and temperature and gravity calibrations 
(while still systematically uncertain) have been significantly refined 
(Luhman et al. 2003; Lyo et al. 2008; Cruz et al. 2009). The radii of 
solar-type stars have been studied in similar detail, but corresponding 
progress for low-mass stars has not maintained the same pace. These radii 
are a crucial component in testing stellar models because small changes in 
opacities and convective efficiencies can significantly change the 
interior structure, leading to significance changes in the expected radius 
at a given mass (e.g. Chabrier et al. 2007).

The sample of low-mass ($\la$0.7 $M_{\sun}$) stellar radii has been 
gradually assembled from many sources over the past $\sim$5 years. The 
best radius measurements tend to come from eclipsing double-lined 
spectroscopic binaries (e.g., Morales et al. 2009; Windmiller et al. 
2010), which can be studied in detail since both components are easily 
observable. Some measurements have fractional uncertainties of $\la$1\%, 
though most systems have not been observed with sufficient resources to 
yield such precision. Many measurements have also come from single-lined 
or marginally double-lined systems consisting of a higher-mass F/G dwarf 
and a low-mass M dwarf. These systems can be more difficult to 
characterize since they are single-lined spectroscopic binaries, but 
basing the analysis on the better-understood properties of solar-type 
stars can allow for sufficient precision ($\sim$2--5\%) to be helpful. 
Finally, the newest technique to yield new radius measurements is 
long-baseline optical interferometry (e.g. Berger et al 2006; Demory et 
al. 2009; Boyajian et al. 2010), which can yield radius measurements for 
single stars and not just close binaries, but is typically limited to only 
a small sample of the closest, brightest stars and faces a systematic 
limit of $\sim$5\% in the determination of masses (Delfosse et al. 2000).

In Figure 10, we show the updated mass-radius relation for our new 
observations, as well as for all previous measurements that have 
fractional uncertainties of $<$3\% and fall in the same mass range 
(0.35--0.65 $M_{\sun}$). We also show the theoretical mass-radius relations 
for old (1 Gyr and 5 Gyr) field stars as predicted by the models of 
Baraffe et al. (1998). The components in four of our newly-discovered 
systems sit very close to the theoretical mass-radius relation, which 
seems like an encouraging endorsement for the models. However, one or both 
components for our other two systems sit significantly above the model 
sequence, as do most measurements obtained in previous studies; this trend 
has been well-known for several years (e.g. Lopez-Morales 2007). We do not 
expect a dispersion this large from systematic effects (Section 5.4), so 
it seems plausible that the variations in observed stellar properties 
could be genuinely astrophysical in origin, perhaps due to variations in 
stellar activity and magnetic fields that change the convective efficiency 
in stellar envelopes (e.g. Chabrier et al. 2007).

Close binaries are known to be significantly more active than wide 
binaries and single stars (e.g. Shkolnik et al. 2010), most likely due to 
their tidally-locked high rotational velocities. Most of the known 
low-mass eclipsing binaries show H$\alpha$ in emission, including all of 
our newly-discovered systems (Table 8), while typical early-M stars show 
significant H$\alpha$ emission only within $\la$1 Gyr after formation 
(West et al. 2008). If this rotation-driven activity is the root cause for 
MDEBs' inflated radii, then we might expect longer-period systems (with 
correspondingly lower rotational velocities) to show a smaller effect, and 
eventually to reach the same mass-radius relation as for single stars. 
This hypothesis has been difficult to test because the vast majority of 
known systems have very short periods ($\la$0.5--1.0 d), and the few known 
long-period systems have not been characterized to high precision. 
However, our sample includes several systems with periods of 1.5--2.0 d. 
We therefore can test for a difference in the radii of ``short-period'' 
and ``long-period'' systems, where the sample is divided to yield two 
equally-sized samples. There are no precisely-characterized systems with 
periods of 0.85--1.5 days, so the division could be placed anywhere in this 
range.

In Figure 11, we plot the model-normalized radius ($R_{obs}/R_{model}$) as 
a function of orbital period for all of the known eclipsing binary systems 
that have component masses of 0.35--0.80 $M_{\sun}$ and fractional 
uncertainties of $<$3\% in their masses and radii. Each measurement has an 
uncertainty in $R_{obs}/R_{model}$ that encompasses the direct error in 
$R_{obs}$ as well as the implicit error in $R_{model}$ that comes from the 
uncertainty in $M_{obs}$; the uncertainties in period are negligible on 
this scale.  Short-period systems ($P\la$1 day) are systematically larger 
($+4.8 \pm 1.0$ \%, with standard deviation of 3.4\%) in 
$R_{obs}/R_{model}$, with some radii up to 10\% larger than the models. In 
contrast, most systems (including all of our new systems) with periods of 
$\sim$1.5--2.0 days show much better agreement with theoretical 
predictions ($+1.7 \pm 0.7$ \%, with standard deviation of 2.4\%), and 
only 1/12 is $>$5\% larger than model predictions. The mean radii for the 
two populations are therefore distinct with a significance of 2.6$\sigma$. 
It is also plausible that the dispersion for long-period systems could 
result from differences in analysis techniques; our four systems alone 
have a dispersion of only $\pm$1.8\%. In contrast, most of the dispersion 
for short-period systems can be seen between components of the same 
system. The components of NSVS0103, GU Boo, and NGC 2204-S892 differ by 
$6.0\pm1.3$\%, $3.3\pm1.8$\%, and $5.3\pm3.5$\%, respectively.

This trend seems to confirm that close eclipsing binary systems are indeed 
inflated in comparison to most low-mass stars, and since they are poor 
representatives of typical low-mass stars, then any discrepancies with 
respect to theoretical models should not be taken as an indictment of 
those models. There are no predictions for the functional form of the 
radius-period relation, so detailed analysis will require additional 
theoretical guidance. Current observations do not sample parameter space 
with enough detail to predict an empirical relation, but the flood of new 
MDEBs expected from upcoming surveys (e.g., Dupuy et al. 2009) should 
allow us to address this shortcoming while new theoretical results are in 
development. In particular, the top priority for these programs should be 
to characterize systems with even longer periods. On average, the 12 
components of long-period systems in our sample agree with stellar 
evolutionary models to within $\la$2\%. However, it seems plausible that 
the radius-period relation could continue to decline for longer-period 
systems, such that long-period binaries and single stars are actually 
smaller than models would predict. Additional high-precision measurements 
also will allow us to test theoretical models with higher precision. Our 
results for long-period systems in Figure 11 suggest that the models do 
still underpredict radii by $1.7 \pm 0.7$ \%.

Alternative explanations also must be considered and ruled out. For 
example, Morales et al. (2010) suggested that the radius discrepancies 
might not be a genuine trend, but instead a measurement artifact resulting 
from nonuniform spot coverage on active stars, and particularly from heavy 
spotting on the stellar poles. This hypothesis is difficult to distinguish 
from the convective inefficiency hypothesis since both should result in 
smaller radii (apparent or real) in longer-period systems. We suggest that 
one possible test might be to measure radii as a function of binary impact 
parameters. Grazing-incidence eclipses will occult a higher relative 
fraction of spotted area than central eclipses will occult, resulting in 
larger apparent radii. Our current sample is not large enough to see any 
apparent trend, but future surveys should discover and characterize many 
more systems. Multi-wavelength observations might also serve to test the 
spot hypothesis, as the contrast between photospheres and spots is less 
severe at long wavelengths. Our existing observations do not reveal a 
significant trend between observation wavelength and measured radius 
because the uncertainties in our $V$ and $R$ lightcurves are too large. 
However, this test should be pursued for bright systems like GU Boo, 
ideally with larger-aperture telescopes than have been used in the past.

 \begin{figure*}
 \plotone{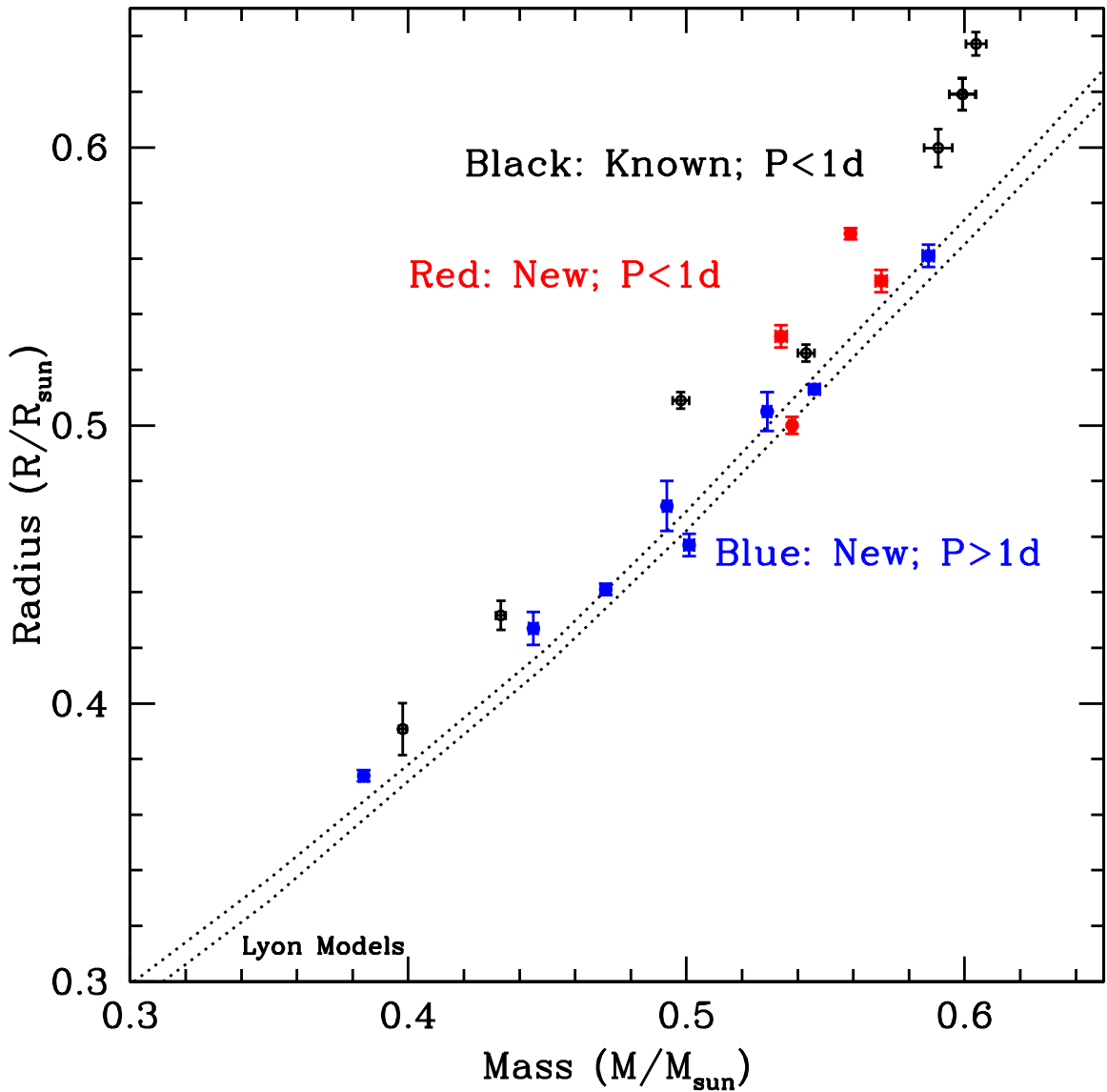}
 \caption{Masses and radii measured for our 12 sample members (filled red: 
$P<1$ day; filled blue: $P>1$ day) and 8 components of known M dwarf 
eclipsing binaries with parameters measured to $\le$3\% precision (open 
black; Torres \& Ribas 2002; Ribas 2003; Lopez-Morales et al. 2006; 
Windmiller et al. 2009). The dotted lines shows the mass-radius relation 
predicted by the low-mass stellar models of Baraffe et al. (1998) at ages 
of 1 Gyr (lower) and 5 Gyr (upper). Many low-mass stars appear to sit well 
above the theoretical mass-radius relation, with some excesses as large as 
10\%. However, the significant scatter observed between stars (and even 
between components of the same system) indicates that an additional factor 
could influence stellar radii. (A color version of this figure is 
available in the online journal.)}
 \end{figure*}

 \begin{figure*}
 \plotone{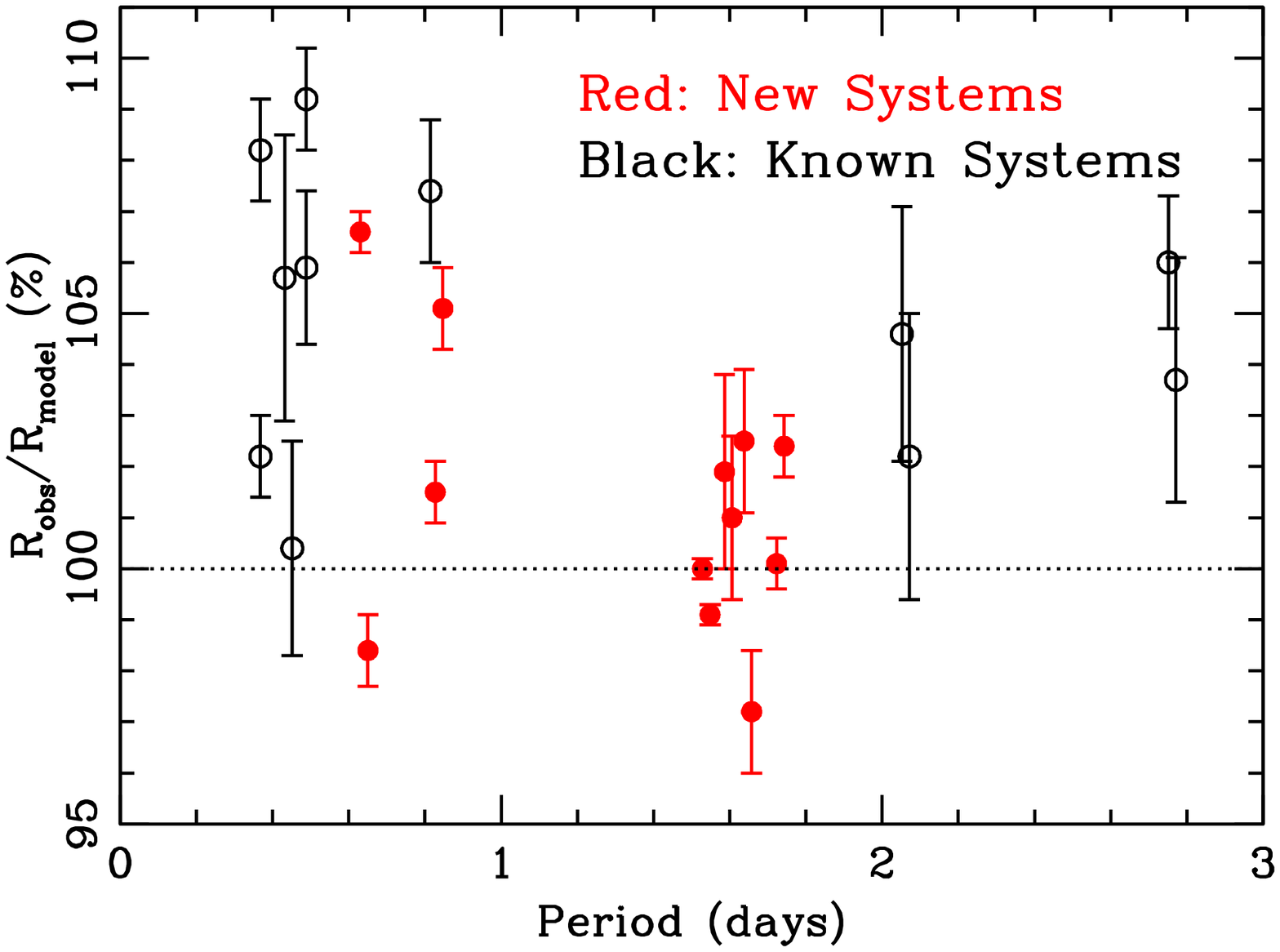}
 \caption{Fractional radius discrepancy ($R_{obs}/R_{model}$) as a 
function of orbital period for our newly-identified eclipsing binary 
systems (filled red) and known systems (open black). In order to isolate a 
possible period dependence, we only use a restricted range of component 
masses (0.35--0.80 $M_{\sun}$) and neglect measurements from the 
literature which are uncertain by $>$3\%, leaving the 12 components of our 
new systems and 12 components of systems from the literature (Torres \& 
Ribas 2002; Ribas 2003; Lopez-Morales et al. 2006; Lopez-Morales \& Shaw 
2007; Windmiller et al. 2009; Rozyczka et al. 2010). Short-period systems 
show a significant spread in radius, such that some components are 
consistent with the models while others are too large by up to 10\%. In 
contrast, the radii of long-period systems tend to consistently be closer 
to model predictions. We suggest that short-period systems tend to be 
inflated because their fast rotation leads to strong magnetic field 
interactions that inhibit convection (e.g. Chabrier et al. 2007); it is 
still unclear why two components in the same system can be inflated by 
different amounts. (A color version of this figure is available in the 
online journal.)}
 \end{figure*}

\section{Summary}

We have discovered and characterized six new M dwarf eclipsing binary 
systems, doubling the number of such systems with well-characterized 
masses and radii ($\sigma \la 3\%$). The components of these systems have 
masses of 0.38--0.59 $M_{\sun}$ and orbital periods of 0.6-1.7 days. The 
shorter-period systems in our sample ($P\la$1 day) tend to follow an 
elevated mass-radius relation that is consistent with the results seen for 
previous systems, most of which also have short periods. The components 
have radii which are up to 10\% larger than are predicted by stellar 
evolutionary models ($\mu=4.8\pm1.0\%$), and the scatter in this relation 
is significantly larger than would be expected from the uncertainties. In 
contrast, longer-period systems have radii that are consistently closer 
to those the models predict ($\mu=1.7\pm0.7\%$).

In light of these results, we conclude that the radii of short-period 
($\la$1 day) eclipsing binaries are most likely inflated in comparison 
to most low-mass stars, and hence they are not good representatives for 
testing stellar evolutionary models. Since these systems show signs of 
high chromospheric activity, including H$\alpha$ emission and significant 
spot coverage, it seems very plausible that the excess radius is a result 
of rotation-driven activity in these tidally-locked rapid rotators, as 
was suggested by Chabrier et al. (2007). Longer-period systems seem much 
more consistent in comparison to each other and the models, so they are 
likely to be better subjects for calibrating the models. However, even 
though these systems are tidally locked into slower rotational periods, 
they still rotate faster than their single brethren, plus they still show 
some signs of heightened activity (including H$\alpha$ emission with 
similar line fluxes as for short-period systems). We suggest that this 
assertion should be tested for systems with even longer periods. Our new 
systems also do not probe the fully convective regime, which could be 
subject to different physics than for stars with convective envelopes and 
radiative cores.

Finally, the high yield of our survey (which doubles the available sample 
of low-mass eclipsing binaries) suggests that a similar strategy (deep, 
sparsely sampled observations over a limited area) could be more rewarding 
than surveys that observe a wider area with shallower limits. As we will 
report in subsequent publications, the MG1 survey of 300 deg$^2$ to a 
limiting magnitude of $R\sim$18 (Kraus et al. 2007) has uncovered at least 
20 new systems with M spectral types, and we have now finished four 
additional surveys. By covering $\sim$4\% of the sky to a depth that is 
$\sim$5 magnitudes deeper, we have achieved a survey volume that is 
$\sim$40 times larger than all-sky surveys like ASAS and ROTSE. In the 
longer term, synoptic all-sky surveys like Pan-STARRS should dwarf even 
our current efforts (e.g. Dupuy et al. 2009), though care must be taken 
not to discover only sources that are too faint for followup.

\acknowledgements

The authors thank C. Slesnick, G. Herczeg, A.M. Cody, M. Ireland, M. Liu, 
T. Dupuy, J. Johnson, M. Lopez-Morales, and K. Stassun for helpful 
discussions and suggestions, as well as Mansi Kasliwal for her assistance 
in scheduling the observations with the Palomar 60" telescope. We also 
thank the anonymous referee for providing a prompt and helpful review. The 
analysis in this paper used two existing data analysis pipelines, MAKEE by 
Tom Barlow and BFall by Slavek Rucinski; we gratefully acknowledge their 
contribution to the field by developing and supporting this software. Some 
of the data products for this project were provided by a collaboration 
between the Global Network of Astronomical Telescopes, Inc. and the Moving 
Object and Transient Event Search System. This work also makes use of data 
products from 2MASS, which is a joint project of the University of 
Massachusetts and IPAC/Caltech, funded by NASA and the NSF. ALK was 
suported by NASA through Hubble Fellowship grant 51257.01 awarded by the 
Space Telescope Science Institute, which is operated by the Association of 
Universities for Research in Astronomy, Inc., for NASA, under contract NAS 
5-26555. This research was partially supported by a grant to GNAT, Inc. 
from the American Astronomical Society.

\clearpage

\begin{deluxetable}{ll|r|r|r}
\tabletypesize{\scriptsize}
\tablewidth{0pt}
\tablecaption{System Properties}
\tablehead{\colhead{Parameter} & \colhead{Units} & 
\colhead{MG1-78457} & \colhead{MG1-116309} & \colhead{MG1-506664}
}
\startdata
\multicolumn{5}{l}{Light Curve Timing}\\
$P$&(days)&1.5862046$\pm$0.0000008&0.8271425$\pm$0.0000004&1.5484492$\pm$0.0000006\\
$T_{0,avg}$&(HJD-2450000)&4758.91630$\pm$0.00010&4783.81137$\pm$0.00003&4573.73166$\pm$0.00003\\
$T_{0,prim}$&(HJD-2450000)&4758.91610$\pm$0.00014&4783.81125$\pm$0.00006&4573.73148$\pm$0.00004\\
$T_{0,sec}$&(HJD-2450000)&4781.91647$\pm$0.00013&4547.66231$\pm$0.00003&4580.69986$\pm$0.00004\\
\hline
\multicolumn{5}{l}{Radial Velocity Orbital Parameters}\\
$\bar{v}$&(km s$^{-1}$)&25.48$\pm$0.09&55.83$\pm$0.15&-13.31$\pm$0.11\\
$K1$&(km s$^{-1}$)&88.41$\pm$0.16&113.31$\pm$0.23&92.33$\pm$0.20\\
$K2$&(km s$^{-1}$)&94.93$\pm$0.16&120.75$\pm$0.23&99.22$\pm$0.20\\
$M_{tot} sin^3(i)$&($M_{\odot}$)&1.0126$\pm$0.0036&1.0986$\pm$0.0045&1.1273$\pm$0.0049\\
$M_A sin^3(i)$&($M_{\odot}$)&0.524$\pm$0.002&0.567$\pm$0.002&0.584$\pm$0.002\\
$M_B sin^3(i)$&($M_{\odot}$)&0.488$\pm$0.001&0.532$\pm$0.002&0.543$\pm$0.002\\
$q$&($M_B/M_A$)&0.931$\pm$0.002&0.938$\pm$0.003&0.931$\pm$0.003\\
$a sin(i)$&($R_{\odot}$)&5.749$\pm$0.007&3.827$\pm$0.005&5.864$\pm$0.009\\
\hline
\multicolumn{5}{l}{Light Curve Fitting Parameters}\\
$i$&(deg)&86.78$\pm$0.05$\pm$0.06&88.74$\pm$0.07$\pm$0.20&88.90$\pm$0.02$\pm$0.09\\
$\Delta T_{eff}$&(K)&64$\pm$6$\pm$8&106$\pm$2$\pm$10&119$\pm$1$\pm$8\\
$T_{eff,A}$\tablenotemark{a}&(K)&3330$\pm$60&3920$\pm$80&3730$\pm$90\\
$T_{eff,B}$\tablenotemark{a}&(K)&3270$\pm$60&3810$\pm$80&3610$\pm$90\\
$R_A$&($R_{\odot}$)&0.505$\pm$0.008$\pm$0.007&0.552$\pm$0.004$\pm$0.013&0.560$\pm$0.001$\pm$0.004\\
$R_B$&($R_{\odot}$)&0.471$\pm$0.009$\pm$0.007&0.532$\pm$0.004$\pm$0.008&0.513$\pm$0.001$\pm$0.008\\
$a$&($R_{\odot}$)&5.758$\pm$0.014&3.828$\pm$0.011&5.865$\pm$0.017\\
$M_A$&($M_{\odot}$)&0.527$\pm$0.002&0.567$\pm$0.002&0.584$\pm$0.002\\
$M_B$&($M_{\odot}$)&0.491$\pm$0.001&0.532$\pm$0.002&0.544$\pm$0.002\\
\multicolumn{5}{l}{Spectroscopic Parameters}\\
EW($H\alpha$)$_{prim}$&($\AA$)&-2.54$\pm$0.08&-0.87$\pm$0.04&-1.12$\pm$0.11\\
EW($H\alpha$)$_{sec}$&($\AA$)&-1.84$\pm$0.05&-1.31$\pm$0.06&-1.01$\pm$0.10\\
 \enddata
 \tablenotetext{a}{Temperature uncertainties are inferred from the 
uncertainty in the SED-fit spectral type (Section 2) and from the 
temperature scale reported in Kraus \& Hillenbrand (2007). Any 
temperatures inferred from such a scale should be regarded as 
systematically uncertain by $\sim$50--100 K. The uncertainty for MG1-646680 
is smaller than the rest because its SED fit includes very precise SDSS 
$ugriz$ magnitudes.}
 \end{deluxetable}

\clearpage

\begin{deluxetable}{ll|r|r|r}
\tabletypesize{\scriptsize}
\tablenum{8}
\tablewidth{0pt}
\tablecaption{System Properties (Cont'd)}
\tablehead{\colhead{Parameter} & \colhead{Units} & 
\colhead{MG1-646680} & \colhead{MG1-1819499} & \colhead{MG1-2056316}
}
\startdata
\multicolumn{5}{l}{Light Curve Timing}\\
$P$&(days)&1.6375302$\pm$0.0000015&0.6303135$\pm$0.0000002&1.7228208$\pm$0.0000042\\
$T_{0,avg}$&(HJD-2450000)&4547.83444$\pm$0.00008&4738.74669$\pm$0.00004&4730.78778$\pm$0.00004\\
$T_{0,prim}$&(HJD-2450000)&4547.83414$\pm$0.00010&4738.74670$\pm$0.00006&4730.78802$\pm$0.00006\\
$T_{0,sec}$&(HJD-2450000)&4579.76658$\pm$0.00013&4739.69215$\pm$0.00004&4755.76843$\pm$0.00006\\
\hline
\multicolumn{5}{l}{Radial Velocity Orbital Parameters}\\
$\bar{v}$&(km s$^{-1}$)&58.77$\pm$0.12&-18.82$\pm$0.08&-3.75$\pm$0.10\\
$K1$&(km s$^{-1}$)&83.25$\pm$0.18&124.79$\pm$0.13&75.43$\pm$0.17\\
$K2$&(km s$^{-1}$)&93.64$\pm$0.18&129.86$\pm$0.14&92.53$\pm$0.18\\
$M_{tot} sin^3(i)$&($M_{\odot}$)&0.939$\pm$0.004&1.0781$\pm$0.0024&0.8456$\pm$0.0038\\
$M_A sin^3(i)$&($M_{\odot}$)&0.497$\pm$0.002&0.550$\pm$0.001&0.466$\pm$0.002\\
$M_B sin^3(i)$&($M_{\odot}$)&0.442$\pm$0.002&0.528$\pm$0.001&0.380$\pm$0.001\\
$q$&($M_B/M_A$)&0.889$\pm$0.003&0.961$\pm$0.002&0.815$\pm$0.003\\
$a sin(i)$&($R_{\odot}$)&5.726$\pm$0.008&3.173$\pm$0.002&5.721$\pm$0.009\\
\hline
\multicolumn{5}{l}{Light Curve Fitting Parameters}\\
$i$&(deg)&87.21$\pm$0.04$\pm$0.07&84.77$\pm$0.04$\pm$0.12&86.08$\pm$0.02$\pm$0.05\\
$\Delta T_{eff}$&(K)&104$\pm$6$\pm$14&83$\pm$2$\pm$14&136$\pm$3$\pm$10\\
$T_{eff,A}$\tablenotemark{a}&(K)&3730$\pm$20&3690$\pm$80&3460$\pm$180\\
$T_{eff,B}$\tablenotemark{a}&(K)&3630$\pm$20&3610$\pm$80&3320$\pm$180\\
$R_A$&($R_{\odot}$)&0.457$\pm$0.006$\pm$0.004&0.569$\pm$0.002$\pm$0.023&0.441$\pm$0.002$\pm$0.002\\
$R_B$&($R_{\odot}$)&0.427$\pm$0.006$\pm$0.002&0.500$\pm$0.003$\pm$0.014&0.374$\pm$0.002$\pm$0.002\\
$a$&($R_{\odot}$)&5.733$\pm$0.016&3.186$\pm$0.005&5.734$\pm$0.017\\
$M_A$&($M_{\odot}$)&0.499$\pm$0.002&0.557$\pm$0.001&0.469$\pm$0.002\\
$M_B$&($M_{\odot}$)&0.443$\pm$0.002&0.535$\pm$0.001&0.382$\pm$0.001\\
\multicolumn{5}{l}{Spectroscopic Parameters}\\
EW($H\alpha$)$_{prim}$&($\AA$)&-1.31$\pm$0.24&-1.02$\pm$0.13&-1.77$\pm$0.12\\
EW($H\alpha$)$_{sec}$&($\AA$)&-0.78$\pm$0.08&-1.09$\pm$0.14&-0.86$\pm$0.06\\
 \enddata
 \tablenotetext{a}{Temperature uncertainties are inferred from the 
uncertainty in the SED-fit spectral type (Section 2) and from the 
temperature scale reported in Kraus \& Hillenbrand (2007). Any 
temperatures inferred from such a scale should be regarded as 
systematically uncertain by $\sim$50--100 K. The uncertainty for MG1-646680 
is smaller than the rest because its SED fit includes very precise SDSS 
$ugriz$ magnitudes.}
 \end{deluxetable}

\end{document}